\documentclass[usegraphicx]{emulateapj}

\begin{document}


\title{Fourier Dissection of Early-Type Galaxy Bars}


\author{R. Buta\altaffilmark{1}, E. Laurikainen\altaffilmark{2}, H. Salo\altaffilmark{2}, D. L. Block\altaffilmark{3}, and J. H. Knapen\altaffilmark{4}}
\altaffiltext{1}{Department of Physics and Astronomy, University of Alabama, Box 870324, Tuscaloosa, AL 35487}
\altaffiltext{2}{Division of Astronomy, Department of Physical Sciences, University
of Oulu, Oulu, FIN-90014, Finland}
\altaffiltext{3}{School of Computational and Applied Mathematics, University of
the Witwatersrand, Johannesburg, South Africa}
\altaffiltext{4}{Centre for Astrophysics Research, University of
Hertfordshire, Hatfield AL10 9AB, UK}



\begin{abstract}
This paper reports
on a near-infrared survey of early-type galaxies designed to provide
information on bar strengths, bulges, disks, and bar parameters in
a statistically well-defined sample of S0-Sa galaxies. 
Early-type galaxies have the advantage that their bars are relatively
free of the effects of dust, star formation, and spiral structure
that complicate bar studies in later type galaxies. We describe the survey 
and present results on detailed analysis of the relative Fourier intensity 
amplitudes of bars in 26 early-type galaxies. We also
evaluate the {\it symmetry assumption} of these amplitudes with radius,
used recently for bar-spiral separation in later-type galaxies.

The results show a wide variety of radial Fourier profiles of bars, ranging
from simple symmetric profiles that can be represented in terms of
a single gaussian component, to both symmetric and asymmetric profiles
that can be represented by two overlapping gaussian components. More
complicated profiles than these are also found, often due to
multiple bar-like features including extended ovals or lenses.
Based on the gravitational bar torque indicator $Q_b$,
double-gaussian bars are stronger on average than single-gaussian
bars, at least for our small sample. We show that published
numerical simulations where the bar transfers a large amount of angular
momentum to the halo can account for many of the observed
profiles. The range of possibilities encountered in models
seems well-represented in the observed systems.

\end{abstract}


\keywords{galaxies: spiral;  galaxies: photometry; galaxies: kinematics
and dynamics; galaxies: structure}


\section{Introduction}

S0 galaxies were introduced to the Hubble sequence by Hubble (1936) as
a means of bridging the apparently catastrophic gap between E7 and Sa galaxies.
After real examples were discovered (see Sandage 1961), the hallmark of
the class became a disk shape (definitely ``later than'' E6 or E7)
and an absence of spiral arms or star formation. SB0 galaxies were originally 
classified as 
SBa by Hubble (1926) even though they also lacked arms. This inconsistency
was corrected in Sandage (1961).

Barred S0 galaxies are extremely interesting because they help to take some 
of the mystery out of S0s in general: a bar is usually a disk feature that 
is closely related to spiral structure in many galaxies 
(e.g., Kormendy and Norman 1979). In addition,
ring features are directly related to bars in spirals, and SB0s may show 
vestiges of similar rings. One could 
therefore ask whether bar properties in S0s might provide any 
clues as to how S0s and spirals might be related in an evolutionary sense.

We are interested in the distribution of bar strengths in S0 galaxies,
a property of very early-type galaxies that has not yet been
tapped for what it might tell us about the evolutionary history of
S0 galaxies. Early-type galaxies have a remarkable array of bar 
morphologies whose Fourier and other properties are worth characterizing in more detail. 
The advantage we have for examining these issues is that early-type galaxies 
show their bars largely unaffected by dust, star formation, and spiral
structure. We can look for subtle structural differences between
different bars and isolate possible different bar types.

We are also interested in comparing early-type barred galaxies with 
the models of Athanassoula (2003, 2005), who demonstrated
the importance of angular momentum transfer to the halo as a means
of producing strong bars. The Fourier profile information we provide here 
is ideally suited to comparison with $n$-body models, and for 
evaluating the symmetry assumption used for bar-spiral separation
(Buta, Block, \& Knapen 2003).

The Near-InfraRed S0 Survey (NIRS0S) is an attempt to obtain
a statistically well-defined database of images of S0s from which the 
properties of S0 bars may be fairly compared to those of spirals.
The Ohio State University Bright Galaxy Survey (OSUBGS, Eskridge et
al. 2002) provides a valuable dataset for studying the properties
of spiral galaxies, and has been fully tapped for bar strength studies
by Block et al. (2002), Buta, Laurikainen, \& Salo (2004=BLS04), Laurikainen,
Salo, \& Buta (2004a), Laurikainen et al. (2004b), and Buta et al. (2005). 
The questions we address with the NIRS0S sample are:
(1) how strong do S0 bars get compared to spiral bars? (2) how does the
distribution of bar strengths in S0 galaxies compare with that for
spirals? (3) what characterizes the morphology of bars in S0 galaxies?
and (4) what are the near-IR luminosity ratios and profile
characteristics of bulges in S0 galaxies?

In describing early-type disk galaxies, we will use specific terminology.
An ``ansae" bar is one showing bright enhancements near the ends
(Sandage 1961). A ``regular" bar does not show such enhancements.
Ansae bars are preferentially found in early-type galaxies while
regular bars are preferentially found in later types, although no
statistical study has quantified the difference. A lens is a feature
showing a shallow brightness gradient interior to a sharp edge (Kormendy
1979). An oval is a broad elongation in the light distribution; it
differs from a conventional bar in lacking higher-order Fourier terms.
If intrinsically elongated, a lens can also be an oval. Ovals are
discussed further by Kormendy and Kennicutt (2004).

In this paper, we focus on a set of 26 galaxies from the NIRS0S
having ovals and/or bars, 
and investigate radial profiles of relative Fourier intensity amplitudes.
We seek to examine the diversity in early-type bars according to the 
symmetry of such profiles.

In section 2, we describe first the rationale for the survey
and the sample selection criteria.
In section 3, we describe the observations made with different
instruments and detectors. The study of the relative Fourier intensity
amplitudes is presented in section 5.
A discussion of the results is presented in section 6.

\section{Sample Characteristics and Nature of the Survey}

The sample for the NIRS0S has been drawn from the Third Reference
Catalog of Bright Galaxies (RC3, de Vaucouleurs et al. 1991). We selected
all galaxies in the revised type range $-$3 $\leq$ T $<$ 2 (S0$^-$
to Sa) having $B_T$ $\leq$ 12.5 and logR$_{25}$ $\leq$ 0.35 as targets
for the survey, which amounts to 170 galaxies. However, any sample
of S0 galaxies is bound to have a bias in the sense that some nonbarred S0s
lacking any distinct structure will likely be classified as elliptical
galaxies. This could have serious effects on a bar strength study.
To reduce the bias somewhat, we have supplemented
the sample with 20 more galaxies that are classified as E or E$^+$ in 
RC3, but which are shown by 
Sandage and Bedke (1994=SB94) to be S0 galaxies.
In these cases, we believe the SB94 types to be more accurate.
Related to this issue is the possibility 
that some galaxies classified as S0 in RC3 are actually E galaxies.
This is shown to be the case with two of our ESO sample galaxies (Laurikainen
et al. 2006). In order to identify such cases, we have to rely on the multicomponent
decompositions.

Another potential problem in an early-type disk galaxy sample is
the misclassification of edge-on disks as bars, something which 
could happen if an edge-on disk is immersed in a significant bulge
component or if the disk is a polar ring. Detailed analysis should identify
such cases.

The sample selection criteria do not perfectly match those 
of the OSUBGS because S0s are not as abundant in RC3 as are spirals for the 
OSUBGS selection criteria (mainly $B_T$ $\leq$ 12.0). 
We decided it was more important to try and match 
the {\it sample size} of the OSUBGS rather than the exact selection criteria.
Although the OSUBGS had no restriction on log$R_{25}$, we imposed such
a restriction on our S0 sample because bar strengths cannot be derived
reliably for highly-inclined galaxies. 

The rationale for observing S0s in the near-IR requires some discussion.
Although both S0's and spirals are
highly flattened disk-shaped systems, S0's mostly lack spiral
structure and star formation.  By default 
such systems tend also to be low in dust content. This means that, in
principle, bar strengths in S0s could be derived in optical bands
with little impact due to dust. The reason we chose not to follow such
an approach is that we wanted to make a fair comparison between bars
in S0s and bars in spirals. In optical bands, bars in spirals may be affected by
star formation, leading dust lanes, and even more disorganized dust lanes,
such that the kind of analysis we describe here is less effective
for determining bar properties.
These effects are minimized or eliminated in the near-IR, and thus to
make the comparison with spirals fair, it is essential that the S0s also
be observed in the near-IR. In addition, some S0s do
have dust, and the impact of this dust is minimized in the near-IR passbands.

\begin{figure}
\figurenum{1}
\plotone{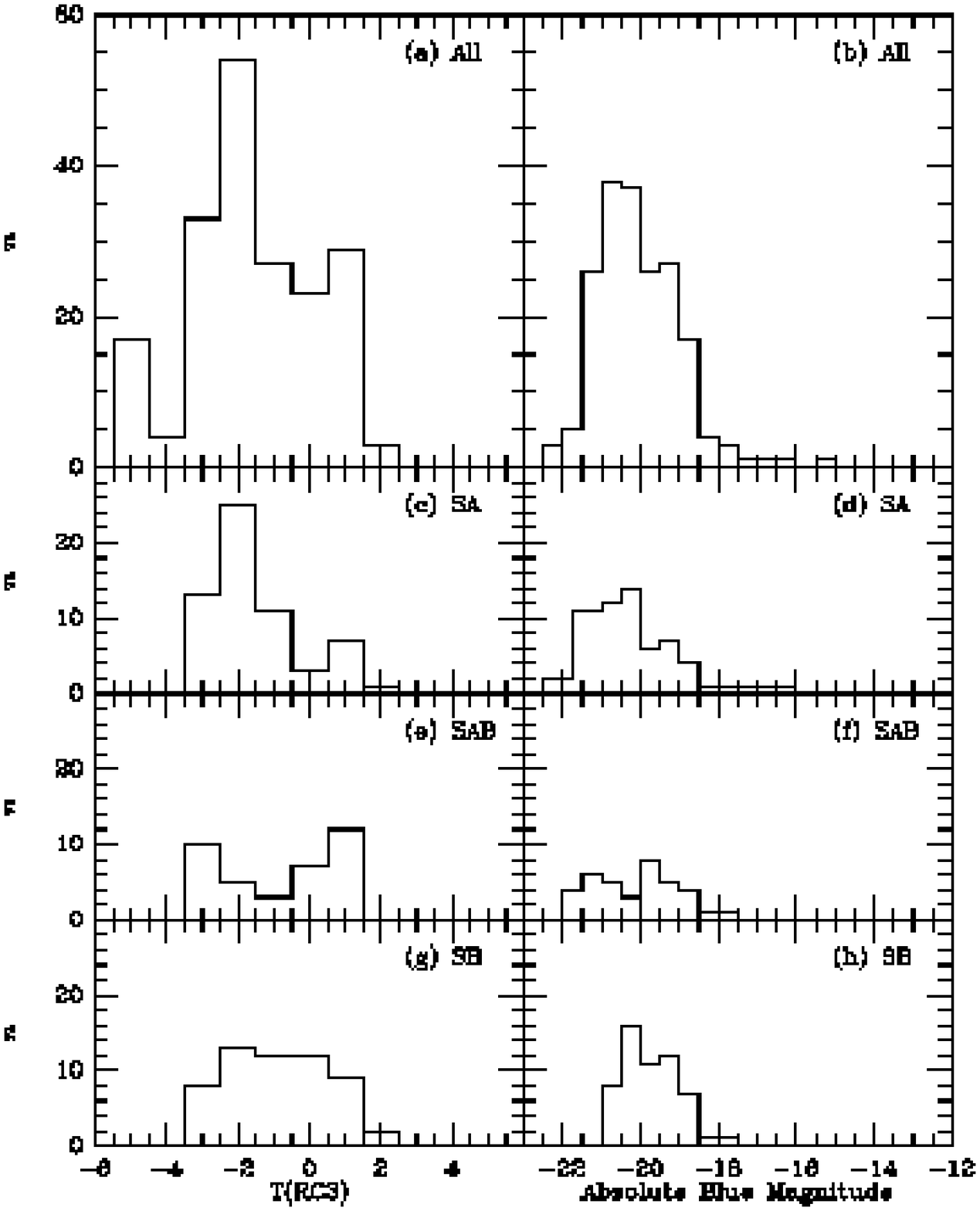}
\caption{}
\label{sample}
\caption{Histograms of numbers of NIRS0S galaxies versus RC3 type index
$T$ and absolute blue magnitude $M_B^o$ based on RC3 data and distances
(mainly from Tully 1988) using a Hubble constant of 75 km s$^{-1}$ Mpc$^{-1}$. The top two
graphs are for the full sample, while the lower ones are divided according
to RC3 family.}
\end{figure}

Figure ~\ref{sample} shows how our sample galaxies are distributed
in type, family, and absolute $B$-band magnitude. Figure ~\ref{sample}a
shows that the dominant type is $T$=$-$2, or S0$^{\circ}$. A significant
fraction of these has no family estimate in RC3, but among the ones
that do, those classified as type SA0$^{\circ}$ contribute the most to
the $T$=$-$2 peak. The unusual distribution for the SAB galaxies in the
sample is partly due to classification uncertainties, as noted by
de Vaucouleurs (1963). SB galaxies in the sample are more uniformly
distributed across the type bins.

Table 1 summarizes mean absolute $B$-band magnitudes and RC3 types for
the same samples as illustrated in Figure~\ref{sample}.
The average absolute magnitude for the full sample is comparable to
that for the spiral sample used by BLS04.
Most noteworthy is
that the SB galaxies in the NIRS0S sample have a fainter average absolute
magnitude than the SA and SAB galaxies, the difference being significant 
at the 2$\sigma$ level. Also, the SAB and SB galaxies
are on average of later type than the SA galaxies (3$\sigma$ level) and for
the full sample, which includes some E galaxies. The sample is still biased
in much the same way as the OSUBGS sample of spirals (BLS04), with few galaxies
having $M_B$ $>$ $-$18. 

\section{Observations}

Observations related to NIRS0S have been taken at the Nordic 2.5m Optical
Telescope (NOT) in January and September 2003 and January 2004,
and with the ESO 3.5-m New Technology Telescope (NTT) in December 2004.
The NOT observations were made with NOTCam, a multi-mode instrument 
that can provide images in the 1-2.5$\mu$m region. We used it to observe
45 galaxies in 
wide field imaging mode, which provided a 4$^{\prime}$$\times$4$^{\prime}$
field of view with 0\rlap{.}$^{\prime\prime}$23 pixels. Flat fielding
was accomplished with twilight frames. All of the NOT galaxies were
observed in the 2.15$\mu$m $K$-short ($K_s$) band, whose shorter wavelength
cutoff than the regular $K$-band reduces some of the thermal background
component. (See Persson et al. 1998 for a discussion of this filter and
its relation to the regular Johnson $K$ filter.) 
In some cases, we also observed an object in the $J$-band 
(1.25$\mu$m) in order to use color gradients to evaluate the
constancy of the stellar mass-to-light ratio used in gravitational
torque analyses (Quillen, Frogel, \& Gonz\'alez 1994). Other aspects of
the data reduction are described by Laurikainen et al. (2005=LSB05, paper 1).

The NTT observations were made with SofI, a wide-field IR spectro-imager
designed for the 1-2.5$\mu$m region. We used SofI to image 15 galaxies
with a field of view of 4\rlap{.}$^{\prime}$9$\times$4\rlap{.}$^{\prime}$9
and a pixel size of 0\rlap{.}$^{\prime\prime}$29. As with NOTCam,
the SofI observations were made with a $K_s$ filter supplemented
in a few cases with a $J$-band filter. The reductions for this dataset
are described by Laurikainen et al. (2006, paper 3).

The galaxies observed up to the time of the present analysis were
selected by observing conditions and the time of year, and should
be a representative subset of the whole NIRS0S. A few galaxies outside
the NIRS0S sample are included in the present data set because
some of our NOT runs were carried out under precarious conditions of wind
and seeing that limited our access to NIRS0S
sample galaxies. In this circumstance, we relaxed our magnitude limit somewhat to
accomodate other accessible objects.

\section{Multi-component Decompositions and Deprojected Images}

LSB05 have analyzed 24 of the NOT galaxies observed
for our survey. A decomposition analysis of the ESO sample is provided
in a separate paper (Laurikainen et al. 2006). In addition to providing new information 
on the bulge and disk properties of S0s, these studies also have provided
deprojected images, constructed using orientation parameters based,
when possible, on deep optical images. The deprojected images are
corrected for bulge deprojection stretch by using the decompositions
to remove the bulge, deprojecting the disk alone, and then adding back the
bulge as a spherical component. The decompositions were essential
because relative Fourier intensity profiles would be affected if
bulge shapes were ignored. With corrected images, we can reliably
analyze these profiles for as much intrinsic information as possible.
LSB05 and Laurikainen et al. (2006) list the orientation parameters
used for our deprojections.

Note that in some cases, the deprojection procedure fails because
the bulge has a flattening intermediate between a sphere and a
highly flattened disk. This is manifested as undercorrected light
along the galaxy minor axis, leaving zones deficient in light that
distort the inner parts of the bar.
For our analysis here, we do not use those galaxies (NGC 1350, 3626, 4340)
in our present sample where these deficient zones appeared to be significant. 
For one of these cases (NGC 1079), we used deprojected
images where the bulge is assumed to be as flat as the disk.
This left some ``deprojection stretch" amplitude in the inner regions
but had little effect on the bar.
Also, those galaxies having little or no bar are not included.
This left 26 galaxies for our Fourier study.

Table 2 summarizes types, diameters, distance moduli (Tully 1988),
and absolute $B$-band magnitudes
of these 26 galaxies. Unlike the types used in Table 1 and Figure~\ref{sample},
these are new classifications taken mostly from the
{\it de Vaucouleurs Atlas of Galaxies} (Buta, Corwin, and Odewahn 2006),
and are based on deep optical CCD images 
with the exceptions of NGC 1440, 2781, 2787, and 3941 whose types
Buta has estimated from photographs in Sandage and Bedke (1994).
The classifications are in a modified version of the de Vaucouleurs
(1959) revised Hubble-Sandage system (Buta 1995). We use these
classifications here because some are more accurate than RC3 classifications.
Several galaxies in Table 2 which are classified as types S0$^o$
or S0$^+$ in RC3 are actually early-type spirals. The average revised
$T$ index of the 26 galaxies on the RC3 numerical scale is $-$0.3, or type S0/a.
Thus, this particular subset of the NIRS0S sample is dominated by early-type
transitional spirals.

\section{Relative Fourier Intensity Profiles of SB0-SBa Bars}

The NIRS0S database as of this writing does not have enough galaxies
to reliably tell us about the distribution of bar strengths in
early-type disk galaxies, but the sample is large enough to allow
us to investigate a simple characteristic of SB0 bars: the behavior of
their relative Fourier intensity amplitudes and phases with radius.
This is an important issue because (1) only for early-type galaxies
can the Fourier properties of bars be studied largely unaffected by
spiral structure; (2) the resulting Fourier profiles have the
potential for allowing us to categorize bars with different Fourier
characteristics; and (3) the same kind of analysis can be applied
to $n$-body model bars (e.g., Athanassoula 2003).

A particular question we ask in this paper is: Do SB0 galaxies support the
symmetry assumption used in bar-spiral separation? That is, do the 
relative Fourier intensity amplitudes of all even $m$ terms decline
radially past a maximum in the same or a similar manner as they rose to that
maximum? Buta, Block, \& Knapen 
(2003=BBK) showed that bars can be effectively separated from spirals 
using this kind of assumption. With the symmetry assumption, bars can
be extrapolated into spiral-dominated regions, and removed from an
image. If one adds the axisymmetric background to the mapped bar image,
then the strength of the bar alone can be estimated without the
influence of spiral torques.
Using this approach, Buta et al. (2005) were able to derive the
distribution of true bar strengths (rather than total nonaxisymmetry
strengths) for nearly 150 galaxies in the OSUBGS.
Although the symmetry assumption had some support in previous optical
studies of barred galaxies (e.g., Ohta, Hamabe, \& Wakamatsu
1990), it can only be 
reliably evaluated if a bar is clean, i.e., not affected by dust
or spiral structure. The NIRS0S is therefore ideal for validating this
approach.

\begin{figure}
\figurenum{2}
\plotone{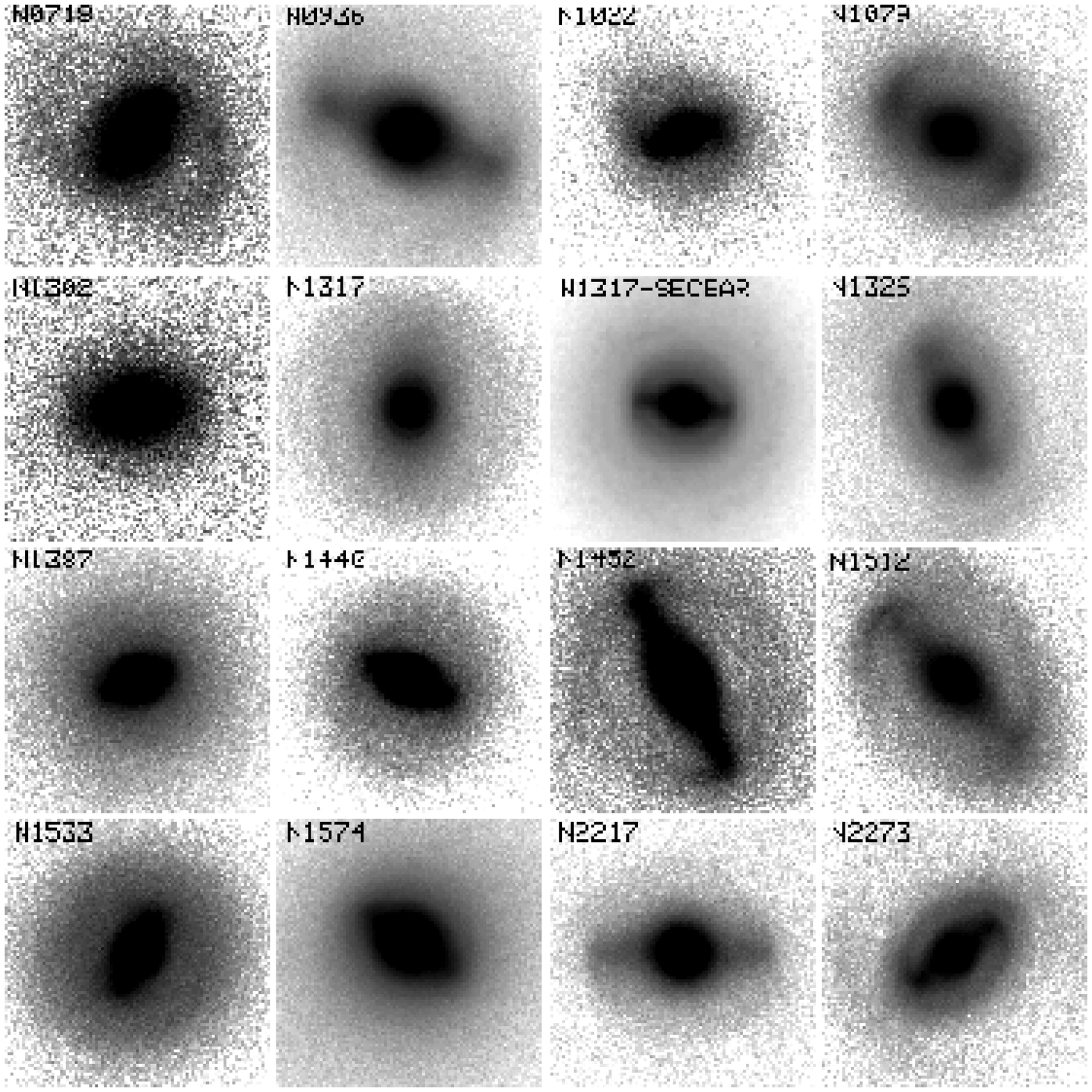}
\caption{}
\label{images}
\end{figure}

Our subset of 26 NIRS0S galaxies includes only those observed so far
having conspicuous
bars or ovals. The deprojected $K_s$-band images of these galaxies are
shown in Figure~\ref{images}. These are displayed mainly to reveal
the primary bars and ovals, but not necessarily secondary bars or ovals,
the discussion of which is covered in more detail in papers 1 and 3 (LSB05; Laurikainen et al. 2006).
We find a variety of Fourier properties of bars. Our main result
is that the relative Fourier intensity profiles of bars in SB0 galaxies
can be represented in simple ways, ranging from symmetric single
component bars, to symmetric multicomponent bars, to asymmetric multicomponent
bars. We discuss the implications of these findings for the bar-halo
interaction scenario proposed by Athanassoula (2003, 2005).

\subsection{Fourier Profiles}

Figure~\ref{afp} shows plots of the relative Fourier intensities 
$I_m/I_0$ and phases $\phi_m$ for our subset of 26 NIRS0S galaxies.
Although we computed these profiles for all even Fourier terms to
$m$=20, Figure~\ref{afp} shows
only the $I_m/I_0$ profiles for  $m$ = 2, 4, 6, and 8 
and $\phi_m$ profiles for $m$ = 2 and 4. The latter
sometimes show sharp changes in phase due merely to the
360$^{\circ}$/$m$ periodicity of the terms.

We can make a
few general observations concerning these profiles. First,
considering that these are near-IR images of early-type
disk galaxies, the profiles of many are surprisingly complex.
Second, more than 2/3 of the galaxies show small peaks
in $m$=2 amplitude near the center that are in some cases
weak nuclear ovals or secondary bars. These
features are almost never aligned with the primary bar
according to the phase plots. Third, the $m$=2 profiles
are often more complicated than the $m$ $>$ 2 profiles,
showing extra components in the nonaxisymmetric 
light distribution.

Interpreting the nuclear features requires some caution.
The Fourier profiles are derived
from deprojected images where the bulge stretch has been corrected,
but only to the extent of approximating the bulge as a spherically
symmetric light distribution. In the final fits of the decompositions
described in LSB05 and Laurikainen et al. (2006), 
most of the bulges were described by elongated light
distributions. Thus, the reliability of the $m$=2 profiles
and phases for the identification of central components
is questionable in some cases. Although we analyze the secondary
bars in NGC 1317 and 3081 in the present paper, we refer to LSB05 and Laurikainen
et al. (2006) for detailed discussions of the other nuclear features. 

Most notable in the Fourier plots are the galaxies whose 
bar profiles are exceptionally symmetric.
Cases like NGC 936, 1326, 1387, 1440,
1533, 1574, 2217, 2787, 2983, 3941, 4596, and 4608
have primary bars that are symmetric enough that 
if these bars coexisted with a spiral, the symmetry 
assumption would recover the bar reliably (section 5.7). 

The remaining galaxies show more complicated profiles with
evidence for multiple components. In some cases, these
components almost do not overlap, as in NGC 1317, 1574,
2681, 2781, and 3081. In each of these cases there is an
inner component whose $I_m/I_0$ profiles are symmetric,
but whose outer profiles may or may not be symmetric.
Other galaxies simply have asymmetric profiles for
a variety of reasons. In the next section, we attempt to
interpret the structure of each bar in the sample.

\subsection{Gaussian Components}

The appearance of many of the profiles in Figure~\ref{afp}
suggests that single or multiple gaussians in fixed position
angles could represent the relative Fourier intensity
profiles of bars. This provides a convenient way of
specifying the radial behavior of bars that could
be useful for comparisons with $n$-body simulations.
Figure~\ref{gaussians} shows our gaussian fits to all
the main profiles in the sample. These were obtained via
nonlinear least squares using 
several routines from {\it Numerical Recipes} (Press et al.
1986). In addition, Figure~\ref{bars}
shows reconstructed images of the mapped bars (excluding the
$m$=0 term), while Figure~\ref{resids} shows the light
remaining after the bar model and the $m$=0 term are
subtracted. The appearance of these plots depends on
how many terms were needed to represent the bar.
Weak bars or ovals required only a few terms, while strong
bars were mapped with all even terms to $m$=20.
The residual images include all odd Fourier terms as
well as any even terms not part of the mapped bar.
The bar mappings use average phases for each term over 
a range of radii.

\subsubsection{Single-Gaussian Fourier Profiles}

Figure~\ref{gaussians} shows that 
some observed bars have $I_m/I_0$ profiles 
that are well-fitted by a single gaussian in all
terms, with little or no amplitude outside this
component. The bars seen in NGC 1387, 1440, 1533, 
2787, and 3941 are like this. Related
are cases like NGC 1574 and 2681, where an
extended oval is both misaligned with and mostly
outside the main bar. In these
cases, the single gaussian characteristic still
applies to the primary bar, which is mostly
separate from the oval. NGC 2217 may also
be in this category, although it has a significant
secondary bar and a faint outer ring, and its
profiles are less regular in appearance.

For these galaxies, the bar plus background disk/bulge
light can be described mathematically as

$$I(r,\phi) = I_0(r)[1+\sum_{m=2}^{4,6,etc.}A_m e^{-{(r-r_m)^2\over 2\sigma_m^2}} \cos m (\phi-\phi_m)] \eqno{1}$$

\noindent
where $A_m$ is a constant, $r_m$ is the mean
radius, $\sigma_m$ is the gaussian width, and
$\phi_m$ is the phase for each Fourier component $m$.
The parameters of these single gaussian representations 
are summarized in Table 3; for NGC 1317 and 3081, the fits
refer to a strong secondary bar.

\subsubsection{Double-Gaussian Fourier Profiles}

Other galaxies in our sample have lower-order $I_m/I_0$ profiles
that are better represented by a double gaussian fit. The
bars seen in NGC 936, 1452, 2983, 4245, 4596, 4608, 
and 4643 show this characteristic. For NGC 936 and 2983, only 
the $m$=2 term required a double-gaussian fit, while for NGC 1452
and 4643, even the higher-order terms required double-gaussian fits.
For these galaxies, the bar plus background disk light
can be described mathematically as  

$$I(r,\phi) = I_0(r)[1+\sum_{n=1}^2\sum_{m=2}^{4,6,etc.}A_{nm} e^{-{(r-r_{nm})^2\over 2\sigma_{nm}^2}} \cos m (\phi-\phi_m)] \eqno{2}$$

\noindent
In principle, $\phi_m$ could be different for the two
gaussian components, but in such a case the two features
would not be part of the same bar and would have to
be analyzed separately.  The parameters of the 
double-gaussian representations are also summarized 
in Table 3, divided into components 1 and 2. 

\subsection{Multi-Component Fourier Profiles}

The remaining galaxies in our sample show more complicated
Fourier profiles compared to the single and double
gaussian cases. The presence of rings, multiple bars,
extended ovals, and sometimes faint spirals is often the
reason why the more simple representations outlined above do
not work. Nevertheless, the profiles in these cases can usually be
partly interpreted in terms of single or double gaussians,
as shown in Figure~\ref{afp}.

For example, in NGC 718 the bar is an unusual
multi-component type including an inner pointy oval,
two bright ansae beyond the ends of this bar, and an extended
oval, all in approximately the same position angle. The double 
gaussian fit shown in Figure~\ref{gaussians} does not include 
the extended oval. Beyond a radius of 35$^{\prime\prime}$,
faint spiral structure also adds amplitude.

In NGC 1022, the bar shows considerable isophote twisting.
We can identify at least two components in this bar in different
position angles, and show an approximate separation in 
Figure~\ref{gaussians} based only on the symmetry assumption
and not on gaussian fits. There is also an outer ring and
weak spiral structure outside the bar region.

The primary feature in NGC 1302 is a pointy oval that can be characterized
in terms of a double gaussian with the two components in slightly
different position angles. There is extra $m$=2 and 4 amplitude
outside this bar that is very faint and difficult to interpret.

NGC 1317 is an interesting case where the primary feature is an
oval, while the secondary bar is a normal bar in a
position angle offset by 90$^{\circ}$.
The primary oval shows a relatively symmetric 
multi-component $m$=2 profile, while the secondary bar
is approximately a single-gaussian type that does not
appear to overlap the primary oval. The two components
show Fourier profiles that are completely separated.

All of the amplitude from 10$^{\prime\prime}$--80$^{\prime\prime}$
has virtually the same phase in NGC 1326.
Its nonaxisymmetry is characterized by an ansae type primary bar imbedded
within a highly elongated oval. The Fourier profiles can be
represented well with a double gaussian fit to the $m$=2, 4, and 6
terms. The asymmetry in each of these terms changes because the
outer Fourier component weakens relative to the inner one with increasing
$m$. There is extra $m$ = 2 and 4 light that appears to be due to
enhancements in the outer ring along the bar axis.

NGC 1512 is an early-type spiral with a strong primary bar and an
elongated inner pseudoring around the bar. The presence of the
ring means we cannot study the bar profile alone. Figure~\ref{gaussians}
shows how we would map the bar using the symmetry assumption. The
double-humped mapping completely removes the bar from the image.
The remaining amplitude is connected with the oval intrinsic
shape of the inner pseudoring.

NGC 2273 is an early-type spiral having a well-defined
bar, inner spiral pseudoring, and an extended oval, all in 
approximately the same position angle. The double-gaussian
representation shown in Figure~\ref{gaussians} provides a good
approximation to the bar. The excess $m$=2 amplitude beyond $r$=40$^{\prime\prime}$
is due to the innermost of the galaxy's two outer rings. This
ring is intrinsically aligned perpendicular to the bar and oval.

As already noted, the classified bar in NGC 2681 is mostly a
single gaussian type,
but outside this bar is a significantly extended and somewhat
pointy oval in a very different position angle. This oval is a
major feature of the galaxy's nonaxisymmetry distribution.

The $m$=2 profile of NGC 2781 shows an approximately gaussian peak
at $r$=6\rlap{.}$^{\prime\prime}$5 due to a small inner disk
in the center, noted by LSB05. Completely separate from this feature is a broad,
extended oval with a slightly asymmetric profile. This feature can
be well-represented by two displaced gaussians in slightly 
different position angles. The two gaussians are shown separately in 
Figure~\ref{gaussians}.

NGC 2859 is very similar to NGC 2781 in having a very broad $m$=2
hump at larger radii. The broad hump includes
a significant ansae-type bar and an extended oval. Figure~\ref{gaussians}
shows a double-gaussian representation of the two features. Significant
$m>$2 amplitude is found mainly for the inner gaussian component.

NGC 3081 is the most interesting of the multi-component cases. One can identify
three clear bar-like features: a secondary bar, a weak primary
bar, and an extended oval in the form of a bright elliptical
inner ring (Buta \& Purcell 1998). The primary and secondary bars
have nearly the same maximum value of $I_2/I_0$, but the extended
oval ring is clearly the dominant nonaxisymmetric feature. The 
secondary bar is well-represented as a single gaussian type.

\subsection{Summary of Bar Components}

Table 4 summarizes the results of the previous section.
It lists the principal feature seen in the Fourier profiles
and any bar-like features seen at larger radii. The information listed
in column 3 of Table 4 shows that ansae-type bars are as
frequent as normal (``regular") bars in this sample.
Some of the barred properties are highlighted in 
LSB05 and Laurikainen et al. (2006), as well as in other
publications (e.g., Wozniak et al. 1995; Laine et al. 2002;
Erwin 2004). 

The results show that
Fourier analysis well separates bar/oval/lens structures from spiral arms,
with oval/lens structures affecting both double-gaussian and multi-component
profiles.
One of the interesting results from Table 4 is the large fraction
of the sample galaxies that have a bar which is {\it not}
associated with a strong extended oval. Cases like NGC 1387, NGC 1440,
and NGC 3941 present some of the purest bars (in Fourier space at least) known.
Nevertheless, our 2D decompositions give somewhat different views of
the structures seen in some of these galaxies. For example,
LSB05 and Laurikainen et al. (2006) note that including a lens in the decomposition models of
NGC 1387, 1440, 1533, 1574, and 3941 improves the fits (but has no effect on bulge-to-total
luminosity ratios); all of these show single gaussian Fourier
profiles. In some of these cases, the lens
in question is more of an outer lens (Kormendy 1979) than an inner
lens and the features have little impact on the Fourier 
terms because they are not very barlike. In the cases of NGC 1533 and 1574, 
significant $m$=2 amplitude is seen outside the bar region (Figure~\ref{gaussians}).
Fourier decomposition and 2D multi-component decompositions 
do not always detect the same structural components because the
former is designed to characterize non-axisymmetric components, while
the latter must model all components, both axisymmetric and nonaxisymmetric.
The two approaches are complimentary, and we use Fourier analysis here
only to characterize aspects of the bars seen.

In contrast to NGC 1387, 1440, and 3941, many galaxies have bars 
imbedded in more significant extended ovals. 
These ovals are often aligned with the
main bar, but misalignment is also found. Also, the ovals
may lie so far out that they are distinct entities in
their own right. NGC 2859 is an example of an ansae-type
bar imbedded in a considerably extended, aligned oval, while
NGC 1079 is an example of an ansae-type bar in a slightly
misaligned extended oval. It is important to note that the
extended ovals are distinct from their associated bars.
As seen in NGC 2859, the main bar has significant higher-order
Fourier terms, while the extended oval only appears in the $m$=2
term.

The mean Fourier parameters for the objects fitted with single
or double gaussians in Table 3 are summarized in Table 5. The results show that
bars fitted by single gaussians on average are shorter than those 
requiring double gaussians and have a
smaller gaussian width. The single gaussian bars have $<r_2>$ = 1.8 $\pm$ 0.3 kpc
and $<\sigma_2>$ = 0.58 $\pm$ 0.10 kpc, compared to  $<r_2>$ = $<r_{12}>$
= 2.8 $\pm$ 0.4 kpc
and $<\sigma_2>$ = $<\sigma_{12}>$ = 0.88 $\pm$ 0.12 kpc for the double 
gaussian bars. The mean absolute
blue magnitudes are nevertheless very similar between the two groups
(last line of Table 5). Single gaussian bars have smaller relative Fourier
amplitudes of higher order terms, and less increase in $r_m$ with increasing
$m$, as compared to at least the first gaussian component of the
double gaussian bars. The gaussian widths tend to be larger for $m$=2 
than for $m>$2.

\subsection{Elmegreen Bar Classifications}

An important additional piece of information is the classification of these
galaxies according to their bar surface brightness profiles.
Elmegreen and Elmegreen (1985) showed that bars can be divided
into flat and exponential types according to the shapes of these
profiles. Figure~\ref{barprofs} shows surface
brightness profiles along the primary bar major and minor axes for 25
of our sample galaxies. (The profiles shown for NGC 1317 are instead for
the secondary bar.) The profiles are normalized to the radius $r_2$
(or the maximum of $r_{12}$ and $r_{22}$) of the $m$=2 Fourier
term. Also, the surface brightness is shown relative to that at
the radius $r_2$. From these profiles we have derived the classification
of the main bar given in Column 5 of Table 4. The profiles show a wide range of shapes that
includes both of the Elmegreen types, although half of the galaxies
are clearly intermediate between flat and exponential. Only
two of our sample galaxies are of the exponential type, which is not
unexpected given that Elmegreen and Elmegreen (1985) found
exponential profiles mainly for later type galaxies. Table 4
also shows that there is no correlation between Elmegreen
bar type and the Fourier profile characteristics. However,
the ansae-type bars are more often of the flat type than are
the regular bars.

\subsection{Bar Strengths}

Bar strengths provide an additional way of comparing the various
Fourier profile categories among our sample galaxies.
We have used the gravitational torque approach (Sanders \& Tubbs 1980; Combes
\& Sanders 1981; Buta \& Block 2001)
to derive the bar strength, $Q_b$, from our gaussian-fitted or symmetry-mapped
bar images. This parameter is derived assuming a constant mass-to-light
ratio and negligible dark matter, but allows for the less flattened shape of the
bulge and an exponential vertical density distribution having vertical scaleheight
derived from a fraction of the radial scale length. The procedure 
is described by Laurikainen $\&$ Salo (2002), 
Salo et al. (1999, 2004), and Laurikainen, Salo $\&$ Buta (2004).
From the force in the plane, we derive $Q_b$ as the maximum of the function

$$Q_T(r) = { ~~~~~|F_T(r,\phi)|_{max} \over <|F_R(r,\phi)|>},$$

\noindent
where $|F_T(r,\phi)|_{max}$ is the maximum tangential force and $<|F_R(r,\phi)|>$
is the azimuthally-averaged radial force.
The resulting values of $Q_b$ are compiled in Table 6, and differ only slightly from the {\it total}
nonaxisymmetric strength $Q_g$ presented for these galaxies by LSB05
and Laurikainen et al. (2006).

A comparison between Tables 4 and 6 shows that single gaussian bars have a lower average
$Q_b$ than do double gaussian bars. For the 7 single gaussian cases, $<Q_b>$ = 0.116 $\pm$ 0.049 (stan. dev.)
while for 10 double gaussian cases, $<Q_b>$ = 0.222 $\pm$ 0.106 (stan. dev.). For the 9 remaining
cases, $<Q_b>$ = 0.161 $\pm$ 0.068 (stan. dev.) using the main bar feature when
multiple components are listed. The results suggest that single gaussian bars are weaker on average
than double gaussian bars, although the two groups overlap somewhat in $Q_b$.

The ``$Q_b$ family" in Table 6 is a quantitative family estimate
based on $Q_b$ (Buta et al. 2005). The $Q_b$ family is SA if $Q_b$ $<$ 0.05,
S$\underline{\rm A}$B if $Q_b$=0.05-0.10, SAB if $Q_b$=0.10-0.20, 
SA$\underline{\rm B}$ if $Q_b$=0.20-0.25, and SB if $Q_b$ $\geq$ 0.25.
These ranges were chosen by Buta et al. (2005) to approximate the
way these classifications are often made, especially for the underline types.
Interestingly, from Table 2, 17 of our galaxies are visually judged to be type SB,
while only 6 are SB from the $Q_b$ family. The bars in these early-type galaxies
are not as strong as they look owing to the significant background bulge and disk
components in most cases.  The strongest bar in our sample is that seen in NGC 1452, with
$Q_b$=0.42. Although this kind of bar strength is not unusual for a late-type
spiral, it is exceptional in our early-type galaxy sample.

\subsection{Evaluation of the Symmetry Assumption}

The symmetry assumption of relative Fourier intensity profiles
was used by Buta, Block, \& Knapen (2003) as a straightforward
way of separating bars from spirals that allows the determination
of quantitative bar strength with the effects of spiral arm
torques largely removed. The assumption was based on the 
appearance of such profiles in the $B$-band for six early-type
barred galaxies (NGC 1398, 2217, 4440, 4643, 4650, and 4665; 
types S0$^+$ to Sab) shown by Ohta, Hamabe, \& Wakamatsu (1990).
Although first suggested by $B$-band imaging, the near-IR is
still a better band for evaluating this issue because of
its reduced sensitivity to dust and star formation.

We have shown that relative Fourier intensity profiles of
bars in early-type galaxies show considerable symmetry
in some cases, and asymmetry in others. Here we evaluate
the impact on $Q_b$ of applying the symmetry assumption to two
early-type bars with asymmetric $I_m/I_0$ profiles.

In our sample, NGC 1452 is a strong case of asymmetry in its
bar profile. Figure~\ref{n1452profs} shows two different
applications of the symmetry assumption to the low order
Fourier terms. In the first, the $m$=2 amplitudes are reflected
around $r_2$=24$^{\prime\prime}$ while in the second these amplitudes
are reflected around $r_2$=30$^{\prime\prime}$. The higher order
terms simply follow these mappings with $r_m$ increasing with $m$.
Figure~\ref{n1452resids} shows the light remaining after these
bar extrapolations are applied. For the $r_2$=24$^{\prime\prime}$
case (left panel), the bar is clearly not fully removed, showing
residual ansae. The $r_2$=30$^{\prime\prime}$ case
does a much better removal of the bar. Still, one can see in the map 
(at the bar ends) where too much bar light has been removed, as expected.
The $r_2$=30$^{\prime\prime}$ mapping is very similar to what we have
used here for NGC 1512. The double-humped representation in that case
removes the observed bar from the $K_s$-band image very well, and
reveals the spiral arms near the ends of the bar.

To evaluate the impact of the two representations of NGC 1452's
bar on bar strength, we have computed $Q_b$ for the double-gaussian
representation in Figure~\ref{gaussians} and the two representations
in Figure~\ref{n1452profs}. In all three cases, we assumed a vertical
scale height of 10\rlap{.}$^{\prime\prime}$0 (1.1 kpc using the distance
from Tully 1988). The double-gaussian fit gives $Q_b$=0.416 (Table 6), while the
$r_2$=24$^{\prime\prime}$ case gives $Q_b$=0.356 (14\% difference) and the 
$r_2$=30$^{\prime\prime}$ case gives $Q_b$=0.432 (4\% difference).
Thus, even for an extreme case like NGC 1452, the uncertainty in $Q_b$
introduced by the symmetry assumption is relatively small. In most
bar-spiral separations, residuals like those seen in Figure~\ref{n1452resids}
do not occur (Buta et al. 2005). 

Next we apply the symmetry assumption to NGC 4245, a case where the main
peak in the bar might be lost if a strong spiral surrounded the bar.
Figure~\ref{n4245} shows the rising portions of the $m$=2, 4, and
6 profiles reflected around $r$=33$^{\prime\prime}$. This mapping
removes most of the bar but leaves two weak enhancements near the
bar ends (not shown). The double gaussian mapping in Figure~\ref{gaussians} gives
$Q_b$=0.180 (Table 6) while the Figure~\ref{n4245} mapping gives
$Q_b$=0.174, a 3\% difference. 

These two cases demonstrate the effectiveness of the
symmetry assumption as a way to separate spiral torques from bars 
and derive true bar strengths, not only for
the majority of galaxies with well-define, symmetric
Fourier profiles, but even for
those where the Fourier profiles are less well-behaved, such as NGC 1452
and 4245.

\section{Discussion}

Interpretation of the results in this paper requires a theory
that accounts not only for bar strength and pattern speed,
but also the varied shapes of bars. Studies by Athanassoula
(2005 and references therein) show that angular momentum exchange is at
the heart of all of these issues. Critical to the properties
of bars is how much angular momentum is transferred to the
halo. The effect depends on the density
of matter in halo resonances and on how cold or hot the resonant
material is. In order to absorb angular momentum a halo must be
``live", as opposed to a rigid halo that cannot interact with other
galaxy components. Cold, live halos can absorb so much angular
momentum that a bar can grow very strong. Weaker bars form in
warmer, smaller halos or rigid halos, while in hot disks, mainly ovals will
form. The extreme effects of a live halo interaction has led to
the possibility that bars might be found in systems lacking a
background disk in the bar region. Gadotti and de Souza (2003)
present two possible cases of this, although LSB05 present
contrary evidence that does not support their claim.

Athanassoula computes relative Fourier mass profiles for her models
that can be compared to the $I_m/I_0$ profiles presented in this
paper. Provided the mass-to-light ratio in the $K_s$ band is
relatively constant, such a comparison should be fair. The problems
of comparing $n$-body models with real bars are summarized by
Athanassoula \& Misiriotis (2002), who discuss the types of bars
that develop in massive halo (MH), massive disk (MD), and intermediate
models. Considering that the models are not of any specific
galaxy, and are only a few of a large number of models actually 
computed, the relative Fourier mass profiles well resemble the
observed profiles for some of our galaxies. The MH models show
strong $m$=2, 4, 6, and 8 components while these are much
weaker for the MD models, with $m$=6 and 8 being in the noise.
The MH profiles most resemble those for NGC 1452, 2983, 4608, and
4643 in our sample. The model bars have a fairly sharp outer
edge, as is seen in these galaxies. 

Athanassoula (2003) shows how the mass of the halo impacts the
Fourier terms. More massive halos lead to stronger bars.
Her most massive halo model (M$\gamma$3) has Fourier profiles similar to 
those of NGC 1452 and 4643, two of the strongest bars in our
sample. Model MH1, with a less massive halo, has profiles
similar to NGC 936 and 2787, while model MH2, with the lowest
halo mass in the three illustrated, shows mainly an oval
bar and has profiles resembling those of NGC 1302 and 2781.

The good agreement between the models and the observations is
further shown by the simulations of
Athanassoula, Lambert, \& Dehnen (2005), who evaluate the effects of
a central mass concentration (CMC) on the evolution of the
bar models in the previous papers. They show that a CMC has
an effect on the higher $m$ terms, weakening their importance
and leading to more $m$=2-dominated bars. Before the introduction
of a CMC in their massive halo (MH) models, the relative Fourier mass
profiles (their Figure 4) strongly resemble those we observe for 
NGC 4643, which has the sharpest outer edge of all the bars in our 
sample. In this model, the mass ratios $A_4/A_2$ = 0.69, $A_6/A_0$ =
0.47, and $A_8/A_0$ = 0.36, are very comparable with the intensity
ratios found
using the first gaussian component of NGC 4643: $I_4/I_2$=0.66, 
$I_6/I_2$=0.45, and
$I_8/I_2$=0.29, respectively (from Table 3). Comparable values are found for 
NGC 1452 and NGC 4608. 
In their MH models with a CMC, the profiles are more symmetric and
the ratios above are considerably reduced. This is as observed
for most of the other galaxies in our sample whose bars are
weaker. Massive disk
(MD) models in this paper also show similar effects, but the
bars are weaker than for the MH models
and the profile asymmetries are less important.

Bureau \& Athanassoula (2005) present three $n$-body models
designed for deducing diagnostics of edge-on bars. These
models are similar to those used by Athanassoula \& Misiriotis
(2002) and 
include a weak bar, an intermediate strength bar, and a strong
bar, with each model bar surrounded by a strong (and 
largely circular) stellar inner ring. The Fourier decompositions
of these barred galaxy models again resemble the galaxies described
above. In the strong bar model, the
outer end of the bar has a sharp edge, similar to what is seen
in NGC 1452 and 4643 in our sample. The strongest circular
inner rings in our sample are seen in NGC 1452 and 4608 (Figure~\ref{images}).
Weaker circular inner rings are seen in NGC 936, 1317, and 2787,
but in several of our galaxies the rings are highly elongated.
Early-type barred galaxies have a wide
range in intrinsic inner ring shapes, as is typical of such features
(Buta 1995).

The relative Fourier intensity profiles of the bars in our
sample could also include some of the effects of bar destruction.
The bars in our sample that include overlapping extended
ovals could be cases where the bar destruction was in
progress at the time the galaxies became relatively deficient in gas.
Bournaud \& Combes (2002) suggested that bar destruction in the
absence of external gas accretion can lead to an oval lens
feature (see also Kormendy 1979). An extreme example of this 
in our sample could be
NGC 2859, where the impact of the oval lens extends considerably
beyond the ends of the bar. Our Fourier analysis is efficient
for separating the lens from the bar in this case, based on the disappearance
of the higher order Fourier modes in the lens.
Such an extended oval is only weakly
seen in NGC 1452, 4608, or 4643, as if these bars were at
their peaks and had not begun to disintegrate. These same three
galaxies also have little or no central oval or secondary bar.

\section{Conclusions}

We have described a new survey designed to study the near-IR properties of
bars in early-type (S0-Sa) disk galaxies. The goal of the survey is to obtain
a statistically useful database that is comparable in size to that of the
OSU survey of bright spiral galaxies, for the principal purpose of deriving the
distribution of bar strengths in S0-Sa galaxies. LSB05
have presented a multicomponent decomposition analysis of 24 galaxies in
our sample obtained thus far. Adding in a number of galaxies observed with the
ESO NTT (Laurikainen et al. 2006), we have examined in this paper the relative
Fourier intensity profiles of the bars in a subset of 26 early-type
galaxies, for which the multi-component decomposition analysis provided
deprojected images.

The results show that a significant fraction (17/26) of the bars 
in our sample have simple relative Fourier intensity profiles that 
can be described in terms of single or double 
gaussians. The single gaussian Fourier profile types represent the 
simplest types of bars. 
The presence of extended ovals, secondary bars and nuclear ovals,
as well as intermediate types of features, complicates many of the
observed profiles. There is an indication in our sample that single
gaussian bars are weaker on average than double gaussian bars.

Although gaussians can represent the $I_m/I_0$ profiles of some bars, the exact
physical significance of such representations is unclear at the moment. 
Nevertheless, numerical bar models reproduce the types of
Fourier profiles we observe in our early-type galaxy sample.
We have shown that 
the profiles of the strongest bars in our sample resemble those 
found for massive halo models of barred galaxies where the angular
momentum exchange between a bar and a massive live halo can
be very effective. The weaker bars in our sample
may show the effects of CMCs or have hotter and smaller halos,
or just have more massive disks, 
than the stronger bars. In any case, it is clear that $I_m/I_0$
profiles provide a fruitful way of comparing models and simulations,
and that further such comparisons may clarify the relative importance of
different effects.

We have shown that early-type galaxy bars support the symmetry
assumption used in bar-spiral separation studies. Even in an
extreme case of very asymmetric $I_m/I_0$ profiles, the use
of the symmetry assumption will likely lead to less than a 10\%
uncertainty in the estimates of relative bar torque strength $Q_b$
(BBK03).

We thank E. Athanassoula and an anonymous referee for helpful
comments on the manuscript.
RB acknowledges the support of NSF Grants AST 02-05143 and AST 05-07140
to the University of Alabama. EL and HS acknowledge the support of
the Academy of Finland and the Magnus Ehrnrooth Foundation. JHK acknowledges
the support of the Leverhulme Trust, and DLB acknowledges the 
support of the Anglo-American Chairman's Fund.


\centerline{REFERENCES}

\noindent
Aaronson, M., Huchra, J., Mould, J., Schechter, P. L., and Tully, R. B.
1982, \apj, 258, 64

\noindent
Athanassoula, E. 2003, \mnras, 341, 1179

\noindent
Athanassoula, E. 2005, Celestial Mechanics and Dynamical Astronomy, 91, 9

\noindent
Athanassoula, E., Lambert, J. C., and Dehnen, W. 2005, \mnras, 363, 496

\noindent
Athanassoula, E. \& Misiriotis, A. 2002, \mnras, 330, 35

\noindent
Block, D. L., Bournaud, F., Combes, F., Puerari, I., \& Buta, R. 2002,
\aap, 394, L35

\noindent
Bournaud, F. \& Combes, F., 2002 \aap, 392, 83

\noindent
Bureau, M. \& Athanassoula, E. 2005, \apj, 626, 159

\noindent
Buta, R. 1990, \apj, 356, 87

\noindent
Buta, R. 1995, \apjs, 96, 39

\noindent
Buta, R., Block, D. L., \& Knapen, J. H. 2003, \aj, 126, 1148 (BBK03)

\noindent
Buta, R., Corwin, H. G., and Odewahn, S. C. 2006, {\it The De Vaucouleurs
Atlas of Galaxies}, Cambridge, Cambridge University Press

\noindent
Buta, R., Laurikainen, E., \& Salo, H. 2004, \aj, 127, 279 (BLS04)

\noindent
Buta, R. and Purcell, G. B. 1998, \aj, 115, 484

\noindent
Buta, R., Vasylyev, S., Salo, H., and Laurikainen, E. 2005, \aj, 130, 506

\noindent
Combes, F. \& Sanders, R. H. 1981, \aap, 96, 164

\noindent
de Vaucouleurs, G. 1959, Handbuch der Physik, 53, 275

\noindent
de Vaucouleurs, G. 1963, \apjs, 8, 31

\noindent de Vaucouleurs, G. et al. 1991, Third Reference Catalog
of Bright Galaxies (New York: Springer) (RC3)

\noindent
Elmegreen, B. G. \& Elmegreen, D. M. 1985, \apj, 288, 438

\noindent
Erwin, P. 2004, \aap, 415, 941

\noindent
Erwin, P., Beltr\'an, J., Graham, A. W., \& Beckman, J. E.
2003, \apj, 597, 929

\noindent
Erwin, P. \& Sparke, L. S. 1999, \apj, 521, L37

\noindent
Erwin, P. \& Sparke, L. S. 2003, \apjs, 146, 299

\noindent
Eskridge, P., Frogel, J. A., Pogge, R. W., et al. 2002, \apjs, 143, 73

\noindent
Gadotti, D. A. and de Souza, R. E. 2003, \apj, 583, 75

\noindent
Hubble, E. 1926, \apj, 64, 321

\noindent
Hubble, E. 1936. {\it The Realm of the Nebulae}, New York, Dover Publications

\noindent
Kormendy, J. 1979, \apj, 227, 714

\noindent
Kormendy, J. and Norman, C. A. 1979, \apj, 233, 539

\noindent
Laine, S., Shlosman, I., Knapen, J.~H., \& Peletier, R.~F. 2002, \apj, 567, 97

\noindent
Laurikainen, E., Salo, H., \& Buta, R. 2004a, \apj, 607, 103 (LSB04)

\noindent
Laurikainen, E., Salo, H., \& Buta, R. 2005, \mnras, 362, 1319 (LSB05)

\noindent
Laurikainen, E., Salo, H., Buta, R., \& Vasylyev, S. 2004b, \mnras, 355, 1251 

\noindent
Laurikainen, E., et al. 2006, in preparation

\noindent
Ohta, K., Hamabe, M., \& Wakamatsu, K. 1990, \apj, 357, 71

\noindent
Persson, S. E., Murphy, D. C., Krzeminski, W., Roth, \& Rieke, M. J. 1998, \aj, 116, 2475

\noindent
Press, W. H., Flannery, B. P., Teukolsky, S. A. 1986, Numerical Recipes. The Art of
Scientific Computing, Cambridge, Cambridge University Press

\noindent
Quillen, A. C., Frogel, J. A., \& Gonz\'alez, R. A. 1994, \apj, 437, 162

\noindent
Sandage, A. R. 1961, The Hubble Atlas of Galaxies, Carnegie Inst. of
Wash. Publ. No. 618

\noindent
Sandage, A. \& Bedke, J. S. 1994, The Carnegie Atlas of Galaxies, Carnegie
Inst. of Wash. Publ. No. 638

\noindent
Sanders, R. H. \& Tubbs, A. D. 1980, \apj, 235, 803

\noindent
Tully, R. B. 1988, Nearby Galaxies Catalogue, Cambridge, Cambridge
University Press

\noindent
Wozniak, H., Friedli, D., Martinet, L., Martin, P., \& Bratschi, P. 1995, \aaps, 111, 115


\begin{deluxetable}{lccccc}
\tabletypesize{\scriptsize}
\tablewidth{0pc}
\tablecaption{Mean Properties of NIRS0S Sample\tablenotemark{a}}
\tablehead{
\colhead{Sample} &
\colhead{$<M_B^o>$} &
\colhead{mean error} &
\colhead{$<T>$ (RC3)} &
\colhead{mean error} &
\colhead{$N$}
} 
\startdata
full sample  & $-$20.044 &  0.078 & $-$1.595 &  0.128 &  190\\
SA galaxies  & $-$20.114 &  0.151 & $-$1.530 &  0.165 &   60\\
SAB galaxies & $-$20.129 &  0.175 & $-$0.846 &  0.272 &   37\\
SB galaxies  & $-$19.735 &  0.094 & $-$0.905 &  0.184 &   56\\
\enddata
\tablenotetext{a}{Based on RC3 data and distances either
from Tully (1988) or estimated using the linear Virgocentric
flow model of Aaronson et al. (1982) and a Hubble constant
of 75 km s$^{-1}$ Mpc$^{-1}$.}
\end{deluxetable}


\begin{deluxetable}{llcccc}
\tabletypesize{\scriptsize}
\tablewidth{0pc}
\tablecaption{Fourier Analysis Sample\tablenotemark{a}}
\tablehead{
\colhead{Galaxy} &
\colhead{Type} &
\colhead{$\log D_o$} &
\colhead{$M_B^o$} &
\colhead{$\mu_o$} &
\colhead{Telescope}
\\
\colhead{1} &
\colhead{2} &
\colhead{3} &
\colhead{4} &
\colhead{5} &
\colhead{6} 
} 
\startdata
NGC \phantom{0}718 & (R$^{\prime}$)SA$\underline{\rm B}$(rs)a                & 1.38 & $-$19.4 & 31.65 & NOT \\
NGC \phantom{0}936 & SB($\underline{\rm r}$s)0$^+$                           & 1.67 & $-$20.8 & 31.14 & NOT \\
NGC 1022 & (R)SB($\underline{\rm r}$s)a pec                                  & 1.38 & $-$19.4 & 31.33 & NOT \\
NGC 1079 & (R$_1$R$_2^{\prime}$)S$\underline{\rm A}$B($\underline{\rm r}$s)a & 1.53 & $-$19.0 & 31.13 & NTT \\
NGC 1302 & (R$_2^{\prime}$)SAB($\underline{\rm r}$s)0/a                      & 1.59 & $-$20.0 & 31.51 & NOT \\
NGC 1317 & SAB(r)a                                                           & 1.44 & $-$19.3 & 31.14 & NTT \\
NGC 1326 & (R$_1$)SAB(r)0/a                                                  & 1.57 & $-$19.9 & 31.14 & NTT \\
NGC 1387 & SB0$^-$                                                           & 1.45 & $-$18.4 & 31.14 & NTT \\
NGC 1440 & SB(s)0$^+$                                                        & 1.34 & $-$18.9 & 31.30 & NOT \\
NGC 1452 & (R$^{\prime}$)SB(r)a                                              & 1.36 & $-$19.2 & 31.59 & NOT \\
NGC 1512 & SB(r)ab                                                           & 1.95 & $-$18.9 & 29.90 & NTT \\
NGC 1533 & (RL)SB0$^{\circ}$                                                 & 1.43 & $-$19.9 & 30.64 & NTT \\
NGC 1574 & SB0$^-$                                                           & 1.52 & $-$20.0 & 30.64 & NTT \\
NGC 2217 & (R)SB($\underline{\rm r}$s)0/a                                    & 1.66 & $-$20.1 & 31.45 & NOT \\
NGC 2273 & (RR)SAB(rs)a                                                      & 1.54 & $-$20.3 & 32.26 & NOT \\
NGC 2681 & (R)S$\underline{\rm A}$B($\underline{\rm r}$s)0/a                 & 1.57 & $-$20.3 & 30.62 & NOT \\
NGC 2781 & (R$_2^{\prime}$)SAB($\underline{\rm r}$s)0/a                      & 1.45 & $-$19.9 & 32.20 & NOT \\
NGC 2787 & SB(r)0$^+$                                                        & 1.48 & $-$19.2 & 30.58 & NOT \\
NGC 2859 & (R)SB(rl)0$^+$                                                    & 1.62 & $-$20.4 & 32.03 & NOT \\
NGC 2983 & (L)SB(s)0$^+$                                                     & 1.38 & $-$19.6 & 32.19 & NOT \\
NGC 3081 & (R$_1$R$_2^{\prime}$)SAB(r)0/a                                    & 1.33 & $-$20.0 & 32.56 & NTT \\
NGC 3941 & SB(s)0$^o$                                                        & 1.52 & $-$20.2 & 31.38 & NOT \\
NGC 4245 & SB(r)0/a                                                          & 1.46 & $-$17.9 & 29.93 & NOT \\
NGC 4596 & SB(rs)0/a                                                         & 1.58 & $-$19.6 & 31.13 & NOT \\
NGC 4608 & SB(r)0/a                                                          & 1.49 & $-$19.2 & 31.13 & NOT \\
NGC 4643 & SB(rl)0/a                                                         & 1.49 & $-$20.6 & 32.05 & NOT \\
\enddata
\tablenotetext{a}{Explanation of columns: (1) galaxy name; (2) new $B$-band morphological classification 
from the {\it de Vaucouleurs Atlas of Galaxies} (Buta, Corwin, and
Odewahn 2006), except for NGC 1440, 2781, 2787, and 3941 whose types are in the same system
but based on inspection of photographic images in Sandage and Bedke (1994); (3) logarithm of
isophotal diameter at $\mu_B$ = 25.00 mag arcsec$^{-2}$, corrected for extinction and inclination
(from RC3); (4) absolute $B$-band magnitude based on RC3 data and (5) the distance modulus from
Tully (1988), which is based on
a Hubble constant of 75 km s$^{-1}$ Mpc$^{-1}$ in conjunction with a linear Virgocentric flow model
(Aaronson et al. 1982); (6) telescope used for near-IR imaging: NOT = Nordic Optical 2.5-m
Telescope; NTT = ESO 3.5-m New Technology Telescope.}
\end{deluxetable}

\clearpage

\begin{deluxetable}{lccccrrrrrrrrc}
\tabletypesize{\scriptsize}
\tablewidth{0pc}
\tablecaption{Gaussian Fourier Components for 24 Early-Type Barred and Oval Galaxies\tablenotemark{a}}
\tablehead{
\colhead{Galaxy} &
\colhead{$A_2$} &
\colhead{$A_4$} &
\colhead{$A_6$} &
\colhead{$A_8$} &
\colhead{$r_2$} &
\colhead{$r_4$} &
\colhead{$r_6$} &
\colhead{$r_8$} &
\colhead{$\sigma_2$} &
\colhead{$\sigma_4$} &
\colhead{$\sigma_6$} &
\colhead{$\sigma_8$} &
\colhead{$r_2/r_o$}
\\
\colhead{1} &
\colhead{2} &
\colhead{3} &
\colhead{4} &
\colhead{5} &
\colhead{6} &
\colhead{7} &
\colhead{8} &
\colhead{9} &
\colhead{10} &
\colhead{11} &
\colhead{12} &
\colhead{13} &
\colhead{14} 
} 
\startdata
N0718-1 &  0.35 &  0.12 &  0.05 &  0.00 &  13.1 &  12.6 &  13.1 &   0.0 &   4.5 &   2.4 &   2.1 &   0.0 &  0.18\nl
N0718-2 &  0.40 &  0.18 &  0.09 &  0.00 &  22.8 &  20.1 &  21.1 &   0.0 &   3.7 &   2.7 &   2.4 &   0.0 &  0.32\nl
N0936-1 &  0.34 &  0.29 &  0.17 &  0.09 &  27.3 &  36.9 &  37.5 &  38.5 &   8.8 &   9.4 &   8.4 &   7.3 &  0.19\nl
N0936-2 &  0.43 &  0.00 &  0.00 &  0.00 &  42.5 &   0.0 &   0.0 &   0.0 &  10.0 &   0.0 &   0.0 &   0.0 &  0.30\nl
N1079   &  0.68 &  0.28 &  0.13 &  0.06 &  42.4 &  40.2 &  39.9 &  39.9 &  14.7 &   8.8 &   6.3 &   4.7 &  0.42\nl
N1302-1 &  0.30 &  0.09 &  0.00 &  0.00 &  18.3 &  20.0 &   0.0 &   0.0 &   5.3 &   4.7 &   0.0 &   0.0 &  0.16\nl
N1302-2 &  0.23 &  0.06 &  0.00 &  0.00 &  30.3 &  27.4 &   0.0 &   0.0 &   5.3 &   2.2 &   0.0 &   0.0 &  0.26\nl
N1317   &  0.36 &  0.18 &  0.11 &  0.06 &   4.7 &   5.1 &   5.2 &   5.3 &   2.1 &   1.6 &   1.3 &   1.3 &  0.06\nl
N1326-1 &  0.08 &  0.09 &  0.03 &  0.04 &  20.1 &  26.5 &  29.9 &  28.2 &   4.1 &   5.6 &   3.3 &   5.5 &  0.18\nl
N1326-2 &  0.64 &  0.22 &  0.09 &  0.03 &  41.9 &  41.5 &  37.1 &  39.0 &  17.3 &  11.6 &  10.8 &   6.4 &  0.38\nl
N1387   &  0.39 &  0.15 &  0.06 &  0.03 &  18.3 &  17.9 &  17.2 &  17.0 &   6.0 &   5.1 &   4.9 &   4.9 &  0.22\nl
N1440   &  0.48 &  0.23 &  0.11 &  0.05 &  17.3 &  18.4 &  19.4 &  19.6 &   6.8 &   4.7 &   3.7 &   4.4 &  0.26\nl
N1452-1 &  0.75 &  0.53 &  0.43 &  0.28 &  20.5 &  26.8 &  31.0 &  31.4 &   7.3 &   7.8 &   8.4 &   7.7 &  0.30\nl
N1452-2 &  0.67 &  0.36 &  0.18 &  0.12 &  38.3 &  41.9 &  43.7 &  41.6 &  10.0 &   7.0 &   4.3 &   7.1 &  0.56\nl
N1533   &  0.42 &  0.17 &  0.07 &  0.03 &  19.1 &  20.1 &  21.7 &  22.2 &   7.6 &   5.9 &   5.9 &   6.2 &  0.24\nl
N1574   &  0.32 &  0.14 &  0.06 &  0.03 &  13.3 &  13.5 &  13.5 &  13.5 &   4.3 &   3.6 &   3.1 &   3.7 &  0.13\nl
N2217   &  0.51 &  0.25 &  0.15 &  0.09 &  33.4 &  32.9 &  34.0 &  34.5 &  11.7 &   9.4 &   7.4 &   7.5 &  0.24\nl
N2273-1 &  0.20 &  0.23 &  0.15 &  0.08 &  15.0 &  18.4 &  19.1 &  19.5 &   3.2 &   4.8 &   3.9 &   3.7 &  0.14\nl
N2273-2 &  0.60 &  0.20 &  0.06 &  0.05 &  30.9 &  36.0 &  39.3 &  41.5 &  12.1 &   6.7 &   6.7 &   8.2 &  0.30\nl
N2681   &  0.25 &  0.06 &  0.03 &  0.00 &  14.9 &  14.9 &  14.6 &   0.0 &   5.0 &   3.8 &   3.2 &   0.0 &  0.13\nl
N2781-1 &  0.47 &  0.00 &  0.00 &  0.00 &  36.4 &   0.0 &   0.0 &   0.0 &   8.1 &   0.0 &   0.0 &   0.0 &  0.43\nl
N2781-2 &  0.33 &  0.00 &  0.00 &  0.00 &  54.8 &   0.0 &   0.0 &   0.0 &   8.4 &   0.0 &   0.0 &   0.0 &  0.65\nl
N2787   &  0.52 &  0.21 &  0.11 &  0.06 &  34.3 &  36.3 &  36.6 &  37.3 &  10.2 &   7.9 &   7.2 &   6.2 &  0.38\nl
N2859-1 &  0.56 &  0.22 &  0.11 &  0.06 &  40.2 &  39.6 &  40.2 &  40.6 &  13.5 &  11.7 &   9.1 &   8.7 &  0.32\nl
N2859-2 &  0.37 &  0.00 &  0.00 &  0.00 &  66.1 &   0.0 &   0.0 &   0.0 &  11.6 &   0.0 &   0.0 &   0.0 &  0.53\nl
N2983-1 &  0.32 &  0.44 &  0.27 &  0.17 &  14.9 &  25.4 &  25.9 &  26.4 &   4.7 &   7.3 &   5.9 &   5.4 &  0.21\nl
N2983-2 &  0.74 &  0.00 &  0.00 &  0.00 &  25.4 &   0.0 &   0.0 &   0.0 &   7.6 &   0.0 &   0.0 &   0.0 &  0.35\nl
N3081   &  0.28 &  0.10 &  0.05 &  0.00 &   5.4 &   5.3 &   5.2 &   0.0 &   2.0 &   1.6 &   1.4 &   0.0 &  0.08\nl
N3941   &  0.33 &  0.14 &  0.05 &  0.00 &  20.4 &  20.8 &  20.3 &   0.0 &   5.0 &   3.4 &   2.6 &   0.0 &  0.21\nl
N4245-1 &  0.28 &  0.11 &  0.08 &  0.05 &  19.7 &  23.0 &  26.3 &  29.6 &   6.8 &   5.3 &   5.6 &   4.4 &  0.23\nl
N4245-2 &  0.49 &  0.17 &  0.09 &  0.04 &  38.4 &  38.1 &  38.5 &  38.2 &   8.7 &   8.4 &   4.2 &   2.5 &  0.44\nl
N4596-1 &  0.66 &  0.26 &  0.11 &  0.09 &  38.6 &  39.8 &  39.2 &  44.5 &  14.1 &   8.8 &   6.9 &   8.4 &  0.34\nl
N4596-2 &  0.45 &  0.30 &  0.22 &  0.12 &  61.0 &  57.4 &  53.8 &  56.6 &   8.1 &  10.4 &   8.6 &   6.9 &  0.54\nl
N4608-1 &  0.62 &  0.40 &  0.32 &  0.22 &  25.9 &  33.1 &  36.4 &  37.7 &  10.8 &  10.1 &   9.5 &   9.3 &  0.28\nl
N4608-2 &  0.42 &  0.12 &  0.00 &  0.00 &  43.3 &  41.7 &   0.0 &   0.0 &  10.1 &   9.5 &   0.0 &   0.0 &  0.47\nl
N4643-1 &  0.85 &  0.56 &  0.38 &  0.25 &  33.3 &  39.3 &  41.0 &  41.8 &  13.2 &  12.1 &  10.8 &  10.0 &  0.36\nl
N4643-2 &  0.38 &  0.14 &  0.10 &  0.09 &  54.6 &  54.7 &  54.2 &  53.8 &   7.9 &   3.5 &   3.2 &   3.8 &  0.59\nl
\enddata
\tablenotetext{a}{Explanations of columns: (1) Galaxy name. If a double gaussian was fitted to the $I_m/I_0$
profiles, the first gaussian is listed as "-1" while the second is "-2". (2-5): gaussian relative amplitudes
$A_m$ (equation 1) for $m$ = 2, 4, 6, and 8. For a double gaussian fit (equation 2), 
$A_{1m}$ is listed on the first line and
$A_{2m}$ is listed on the second line for a given galaxy. (6-9): mean radii $r_m$ (or $r_{1m}$ and $r_{2m}$
for a double gaussian fit), in arcseconds. (10-13): gaussian width $\sigma_m$ (or $\sigma_{1m}$ and $\sigma_{2m}$
for a double gaussian fit), in arcseconds. (14) ratio of radius $r_2$ to the radius $r_o=D_o/2$ of the standard
isophote having $\mu_B$ = 25.00 mag arcsec$^{-2}$.}
\end{deluxetable}

\clearpage

\begin{deluxetable}{lllll}
\tabletypesize{\scriptsize}
\tablewidth{0pc}
\tablecaption{Summary of Primary Components\tablenotemark{a}}
\tablehead{
\colhead{Galaxy} &
\colhead{Fourier} &
\colhead{Primary} &
\colhead{Other} &
\colhead{Main bar} 
\\
\colhead{} &
\colhead{profile type} &
\colhead{feature} &
\colhead{features} &
\colhead{profile type} 
\\
\colhead{1} &
\colhead{2} &
\colhead{3} &
\colhead{4} &
\colhead{5} 
} 
\startdata
NGC  718 & MCFP  & pointy oval & detached ansae, extended aligned oval & intermediate\\
NGC  936 & DGFP & ansae bar   &                               & flat\\ 
NGC 1022 & MCFP  & distorted bar &  oval disk      & intermediate\\
NGC 1079\tablenotemark{b} & MCFP  & ansae bar & extended misaligned oval  & flat\\
NGC 1302 & DGFP & pointy oval  &                               & exponential\\
NGC 1317 & MCFP  & oval        &                             & intermediate (sec. bar)\\
NGC 1326 & MCFP  & ansae bar & extended aligned oval      & flat\\
NGC 1387 & SGFP & regular bar  &                                   & exponential\\
NGC 1440 & SGFP & regular bar  &                                   & intermediate\\
NGC 1452 & DGFP & regular bar  &                                   & flat\\
NGC 1512 & MCFP  & ansae bar & extended aligned oval             & flat\\
NGC 1533 & SGFP & regular bar  & extended misaligned outer oval            & intermediate\\
NGC 1574 & SGFP & regular bar  & extended misaligned outer oval            & intermediate\\
NGC 2217 & SGFP & regular bar  &                                   & flat\\
NGC 2273 & MCFP  & ansae bar    & extended aligned oval                     & flat\\
NGC 2681 & MCFP  & regular bar  & extended misaligned ansae oval            & intermediate\\
NGC 2781 & DGFP & oval         & extended aligned oval                     & intermediate\\
NGC 2787 & SGFP & ansae bar &                          & flat\\
NGC 2859 & DGFP & ansae bar     & inner oval, extended aligned outer oval  & flat\\ 
NGC 2983 & DGFP & ansae bar    &                                   & flat\\
NGC 3081 & MCFP  & ring/oval & aligned intermediate bar                    & flat\\
NGC 3941 & SGFP & ansae bar    &                                   & intermediate\\
NGC 4245 & DGFP & regular bar  & extended aligned oval                     & intermediate\\
NGC 4596 & DGFP & ansae bar    & extended aligned oval                     & intermediate\\
NGC 4608 & DGFP & regular bar  &                                   & intermediate\\
NGC 4643 & DGFP & regular bar  &                                   & intermediate\\
\enddata
\tablenotetext{a}{Explanation of columns: (1) Galaxy name; (2) the Fourier $I_m/I_0$ profile
type based on gaussian fitting, where SGFP = single gaussian Fourier profile, DGFP = double
gaussian Fourier profile, and MCFP = multi-component Fourier profile (section 5.2); (3) morphology of 
primary bar or bar-like feature;
(4) additional non-nuclear bar-like features;
(5) Elmegreen and Elmegreen (1985) primary bar classification}
\tablenotetext{b}{Deprojection uncertainties complicated analysis; inner regions
affected by bulge deprojection stretch.}
\end{deluxetable}

\clearpage

\begin{deluxetable}{lcccccccr}
\tabletypesize{\scriptsize}
\tablewidth{0pc}
\tablecaption{Mean Fourier Parameters\tablenotemark{a}}
\tablehead{
\colhead{Parameter} &
\colhead{mean} &
\colhead{standard} &
\colhead{mean} &
\colhead{n} &
\colhead{mean} &
\colhead{standard} &
\colhead{mean} &
\colhead{n}
\\
\colhead{} &
\colhead{} &
\colhead{deviation} &
\colhead{error} &
\colhead{} &
\colhead{} &
\colhead{deviation} &
\colhead{error} &
\colhead{}
\\
\colhead{1} &
\colhead{2} &
\colhead{3} &
\colhead{4} &
\colhead{5} &
\colhead{6} &
\colhead{7} &
\colhead{8} &
\colhead{9} 
} 
\startdata
                     &        & SGFP Cases &    &       &        & DGFP Cases &    &       \\
$<r_2>$ (kpc)        &   1.76 &   0.74 &   0.28 &      7&   2.68 &   1.31 &   0.41 &     10 \\
$<s_2>$ (kpc)        &   0.58 &   0.26 &   0.10 &      7&   0.88 &   0.38 &   0.12 &     10 \\
$<r_{22}>$ (kpc)     &   .... &   .... &   .... &     ..&   4.39 &   1.96 &   0.62 &     10 \\
$<s_{22}>$ (kpc)     &   .... &   .... &   .... &     ..&   0.84 &   0.31 &   0.10 &     10 \\
$<I_4/I_2>$          &   0.43 &   0.04 &   0.01 &      7&   0.64 &   0.33 &   0.11 &      9 \\
$<I_6/I_2>$          &   0.20 &   0.05 &   0.02 &      7&   0.44 &   0.22 &   0.08 &      8 \\
$<I_8/I_2>$          &   0.10 &   0.04 &   0.01 &      6&   0.28 &   0.14 &   0.05 &      8 \\
$<I_{10}/I_2>$       &   0.06 &   0.03 &   0.01 &      6&   0.19 &   0.11 &   0.04 &      8 \\
$<I_{42}/I_{22}>$    &   .... &   .... &   .... &     ..&   0.41 &   0.16 &   0.07 &      6 \\
$<I_{62}/I_{22}>$    &   .... &   .... &   .... &     ..&   0.30 &   0.14 &   0.07 &      4 \\
$<I_{82}/I_{22}>$    &   .... &   .... &   .... &     ..&   0.19 &   0.08 &   0.04 &      4 \\
$<I_{102}/I_{22}>$   &   .... &   .... &   .... &     ..&   0.11 &   0.07 &   0.04 &      3 \\
$<r_4/r_2>$          &   1.02 &   0.04 &   0.01 &      7&   1.23 &   0.21 &   0.07 &      9 \\
$<r_6/r_2>$          &   1.04 &   0.07 &   0.03 &      7&   1.33 &   0.25 &   0.09 &      8 \\
$<r_8/r_2>$          &   1.06 &   0.08 &   0.03 &      6&   1.39 &   0.24 &   0.08 &      8 \\
$<r_{10}/r_2>$       &   1.12 &   0.14 &   0.06 &      6&   1.44 &   0.25 &   0.09 &      8 \\
$<r_{42}/r_{22}>$    &   .... &   .... &   .... &     ..&   0.98 &   0.06 &   0.03 &      6 \\
$<r_{62}/r_{22}>$    &   .... &   .... &   .... &     ..&   1.00 &   0.11 &   0.05 &      4 \\
$<r_{82}/r_{22}>$    &   .... &   .... &   .... &     ..&   1.00 &   0.07 &   0.03 &      4 \\
$<r_{102}/r_{22}>$   &   .... &   .... &   .... &     ..&   1.04 &   0.11 &   0.07 &      3 \\
$<s_4/s_2>$          &   0.77 &   0.07 &   0.03 &      7&   0.96 &   0.26 &   0.09 &      9 \\
$<s_6/s_2>$          &   0.68 &   0.11 &   0.04 &      7&   0.88 &   0.24 &   0.09 &      8 \\
$<s_8/s_2>$          &   0.74 &   0.12 &   0.05 &      6&   0.81 &   0.20 &   0.07 &      8 \\
$<s_{10}/s_2>$       &   1.05 &   0.36 &   0.15 &      6&   1.18 &   0.28 &   0.10 &      8 \\
$<s_{42}/s_{22}>$    &   .... &   .... &   .... &     ..&   0.79 &   0.33 &   0.14 &      6 \\
$<s_{62}/s_{22}>$    &   .... &   .... &   .... &     ..&   0.60 &   0.31 &   0.16 &      4 \\
$<s_{82}/s_{22}>$    &   .... &   .... &   .... &     ..&   0.58 &   0.25 &   0.12 &      4 \\
$<s_{102}/s_{22}>$   &   .... &   .... &   .... &     ..&   0.92 &   0.44 &   0.26 &      3 \\
$<M_B^o>$            &  $-$19.53 & 0.70 & 0.26  &      7& $-$19.72 & 0.84 &   0.27 &     10 \\
\enddata
\tablenotetext{a}{Explanation of columns: (1) Fourier parameter (linear diameters use the distance
moduli from Table 1); (2)-(5) parameter means, standard deviations, mean errors, and the number of 
objects for
SGFP cases from Table 4; (6)-(9) the same for DGFP cases from Table 4.}
\end{deluxetable}

\clearpage

\begin{deluxetable}{llcc}
\tabletypesize{\scriptsize}
\tablewidth{0pc}
\tablecaption{Bar Strengths and $Q_b$ Families\tablenotemark{a}}
\tablehead{
\colhead{Galaxy} &
\colhead{$Q_b$ Family} &
\colhead{$Q_b$} &
\colhead{mean error} 
\\
\colhead{1} &
\colhead{2} &
\colhead{3} &
\colhead{4} 
} 
\startdata
NGC  718               & SAB                       &      0.124 &      0.000 \cr
NGC  936               & SA$\underline{\rm B}$     &      0.201 &      0.028 \cr
NGC 1022               & SAB                       &      0.142 &      0.016 \cr
NGC 1079               & SA$\underline{\rm B}$     &      0.242 &      0.052 \cr
NGC 1302               & SAB                       &      0.130 &      0.007 \cr
NGC 1317-primary bar   & S$\underline{\rm A}$B     &      0.091 &      0.008 \cr
NGC 1317-secondary bar & S$\underline{\rm A}$B     &      0.086 &      0.006 \cr
NGC 1326               & SAB                       &      0.161 &      0.014 \cr
NGC 1387               & S$\underline{\rm A}$B     &      0.065 &      0.002 \cr
NGC 1440               & SAB                       &      0.141 &      0.004 \cr
NGC 1452               & SB                        &      0.416 &      0.035 \cr
NGC 1512               & SB                        &      0.270 &      0.015 \cr
NGC 1533               & SAB                       &      0.107 &      0.002 \cr
NGC 1574               & S$\underline{\rm A}$B     &      0.064 &      0.006 \cr
NGC 2217               & SAB                       &      0.170 &      0.008 \cr
NGC 2273               & SA$\underline{\rm B}$     &      0.209 &      0.001 \cr
NGC 2681-inner bar     & SA                        &      0.039 &      0.003 \cr
NGC 2681-outer oval    & S$\underline{\rm A}$B     &      0.061 &      0.004 \cr
NGC 2781               & S$\underline{\rm A}$B     &      0.066 &      0.001 \cr
NGC 2787               & SAB                       &      0.182 &      0.051 \cr
NGC 2859               & SAB                       &      0.105 &      0.004 \cr
NGC 2983               & SB                        &      0.297 &      0.007 \cr
NGC 3081-primary bar   & S$\underline{\rm A}$B     &      0.069 &      0.001 \cr
NGC 3081-primary oval  & SAB                       &      0.153 &      0.013 \cr
NGC 3081-secondary bar & S$\underline{\rm A}$B     &      0.066 &      0.004 \cr
NGC 3941               & S$\underline{\rm A}$B     &      0.085 &      0.005 \cr
NGC 4245               & SAB                       &      0.180 &      0.002 \cr
NGC 4596               & SB                        &      0.271 &      0.061 \cr
NGC 4608               & SB                        &      0.252 &      0.005 \cr
NGC 4643               & SB                        &      0.299 &      0.002 \cr
\enddata
\tablenotetext{a}{Explanation of columns: (1) Galaxy name; (2) quantitative family estimate
following Buta et al. (2005); (3) maximum relative bar torque based on gaussian and other mappings;
(4) mean error excluding systematic effects.}
\end{deluxetable}


\begin{figure}
\figurenum{2 (cont.)}
\plotone{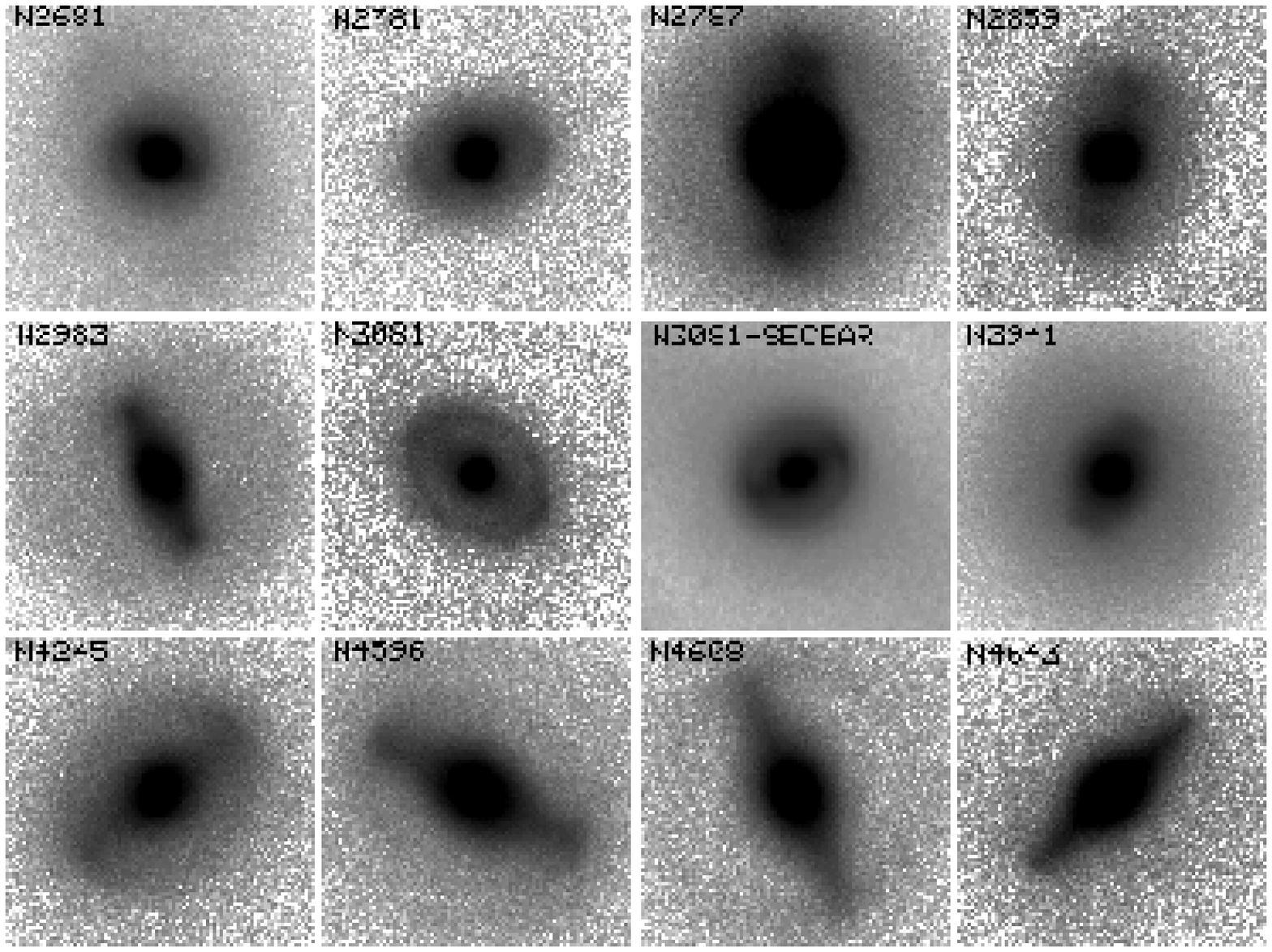}
\caption{Deprojected $K_s$-band images of 26 NIRS0S galaxies. These are
displayed mostly to emphasize the primary bars and outer disks, but not necessarily 
any nuclear structure. The exceptions are the secondary bars of NGC 1317 and
3081, indicated by "SECBAR." The dimensions of each square are 
1\rlap{.}$^{\prime}$96 $\times$ 1\rlap{.}$^{\prime}$96 except for
the following: NGC 1079, 1317, 1326, 1533, 3081 (2\rlap{.}$^{\prime}$47 $\times$
2\rlap{.}$^{\prime}$47);
NGC 2781, 2859, 4596, 4643 (2\rlap{.}$^{\prime}$73 $\times$ 2\rlap{.}$^{\prime}$73); NGC 1512 
(3\rlap{.}$^{\prime}$44 $\times$ 3\rlap{.}$^{\prime}$44); and the secondary bar closeups of NGC 1317, 
3081 (0\rlap{.}$^{\prime}$62 $\times$ 0\rlap{.}$^{\prime}$62).}
\end{figure}

\begin{figure}
\figurenum{3}
\plotone{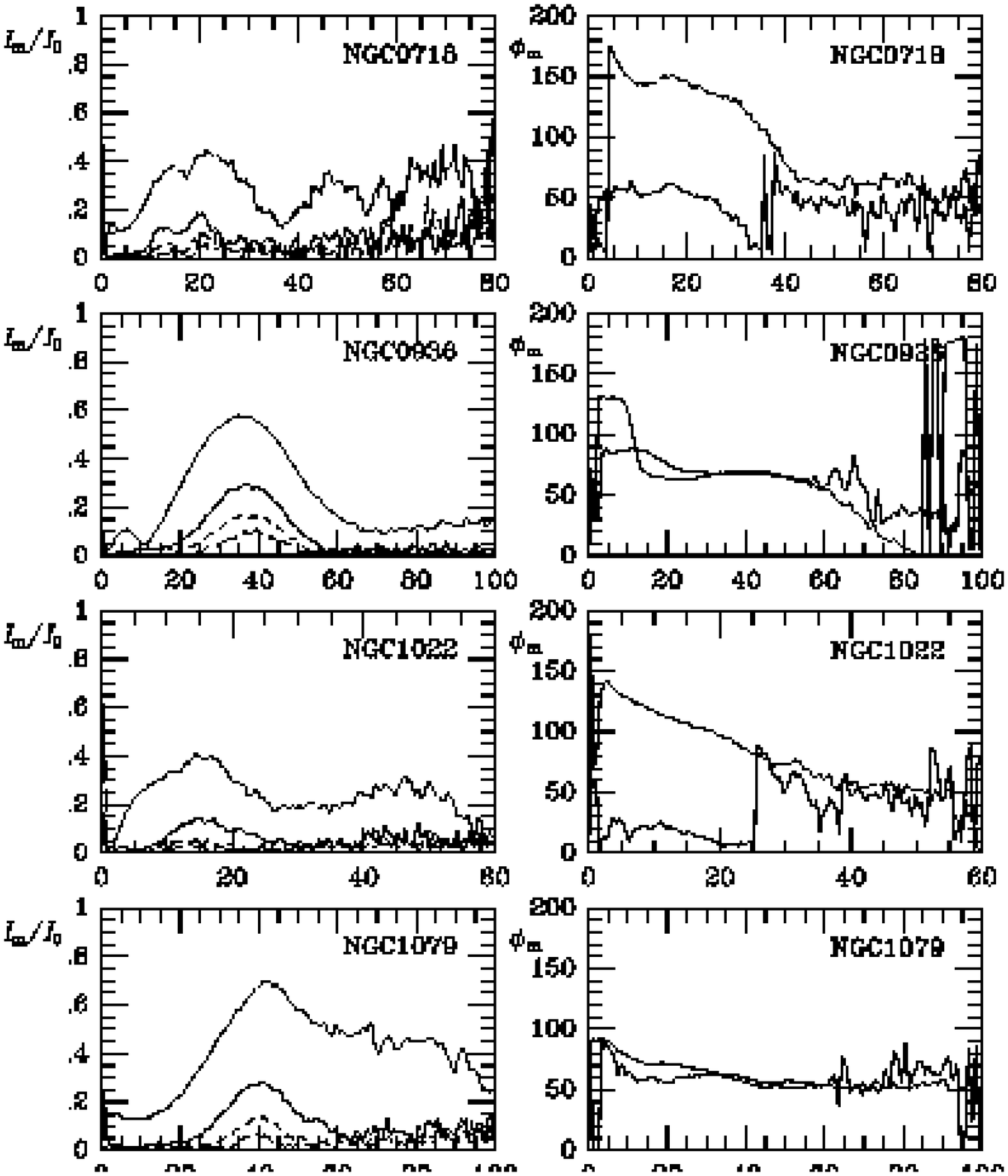}
\caption{}
\label{afp}
\end{figure}
\begin{figure}
\figurenum{3 (cont.)}
\plotone{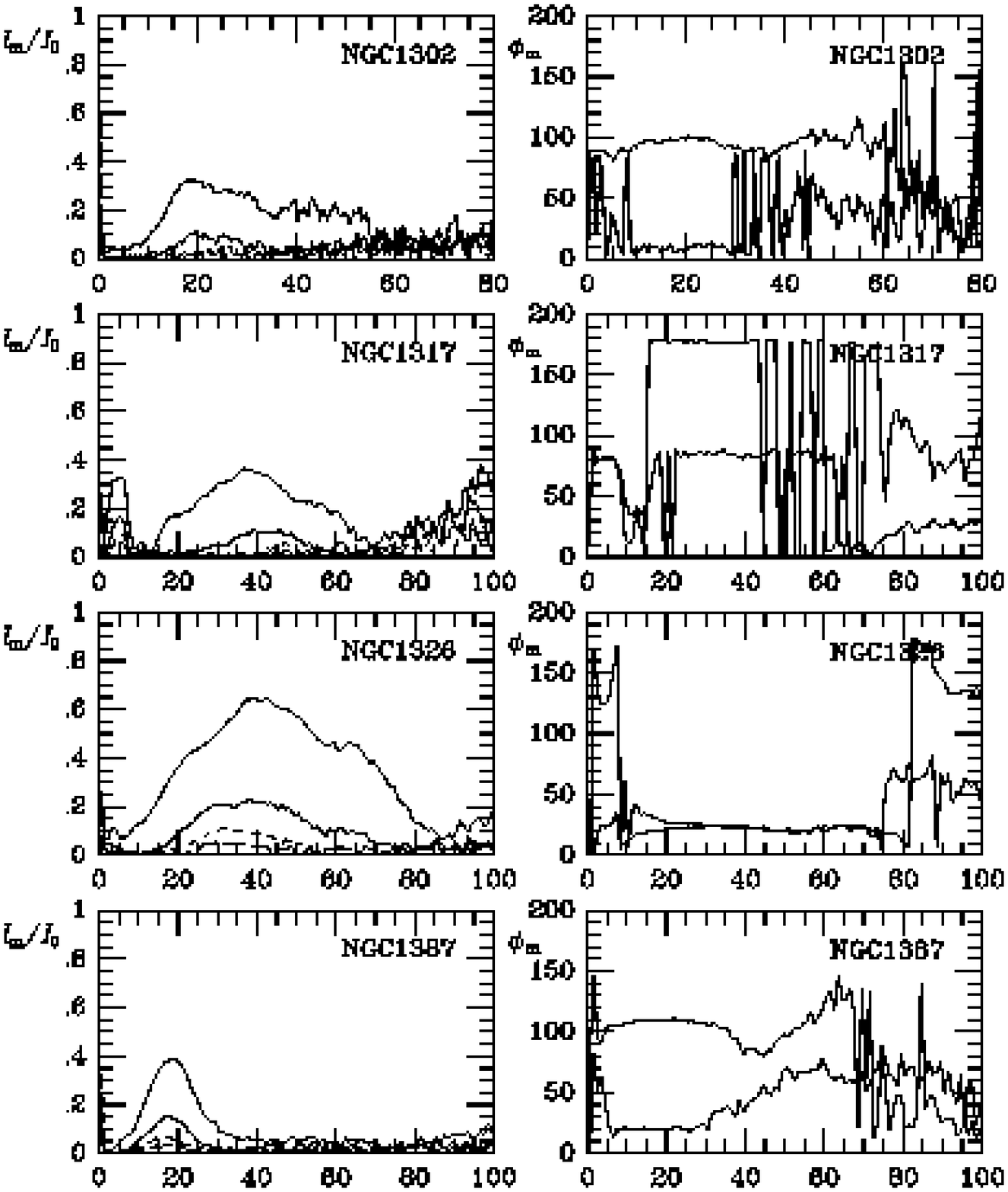}
\caption{}
\end{figure}
\begin{figure}
\figurenum{3 (cont.)}
\plotone{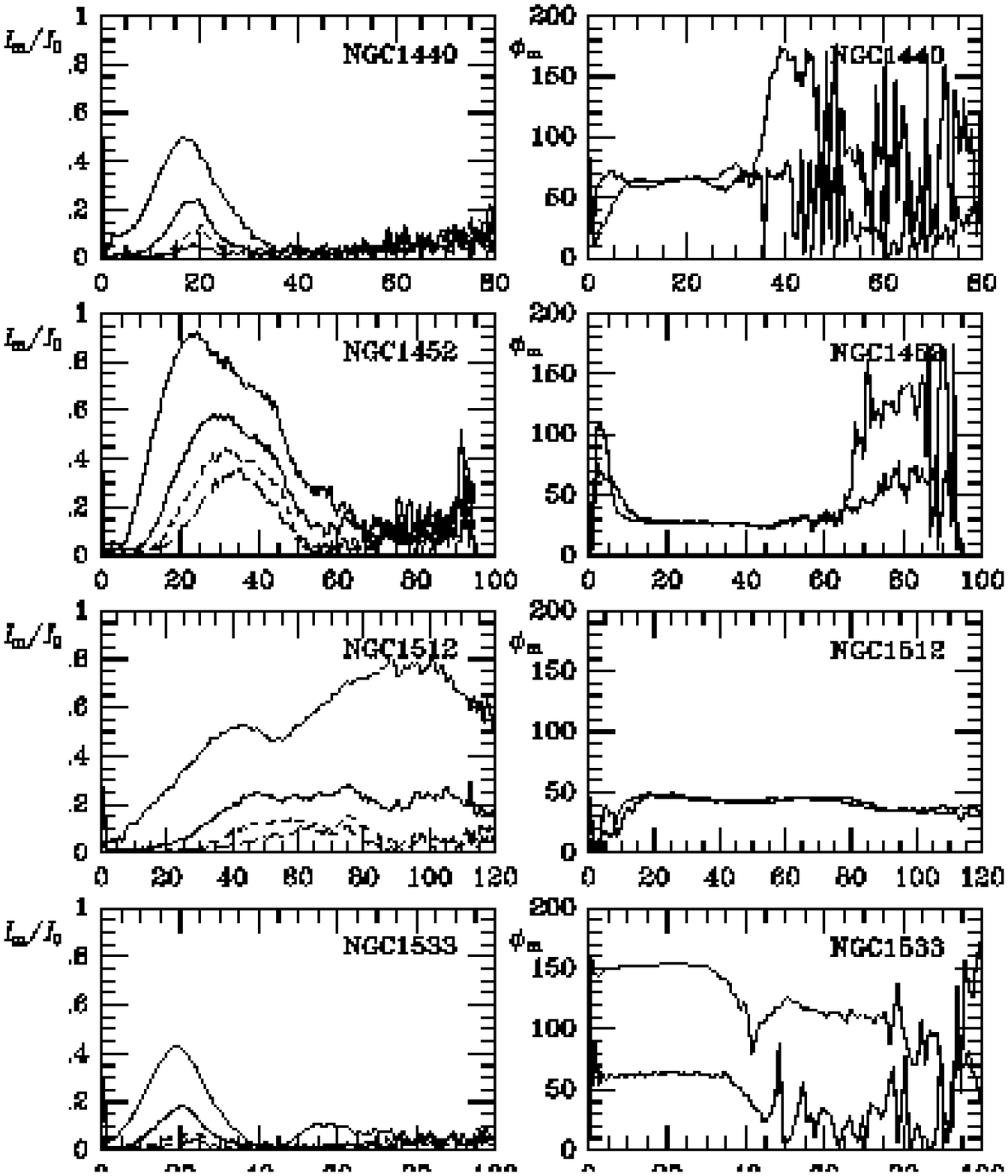}
\caption{}
\end{figure}
\begin{figure}
\figurenum{3 (cont.)}
\plotone{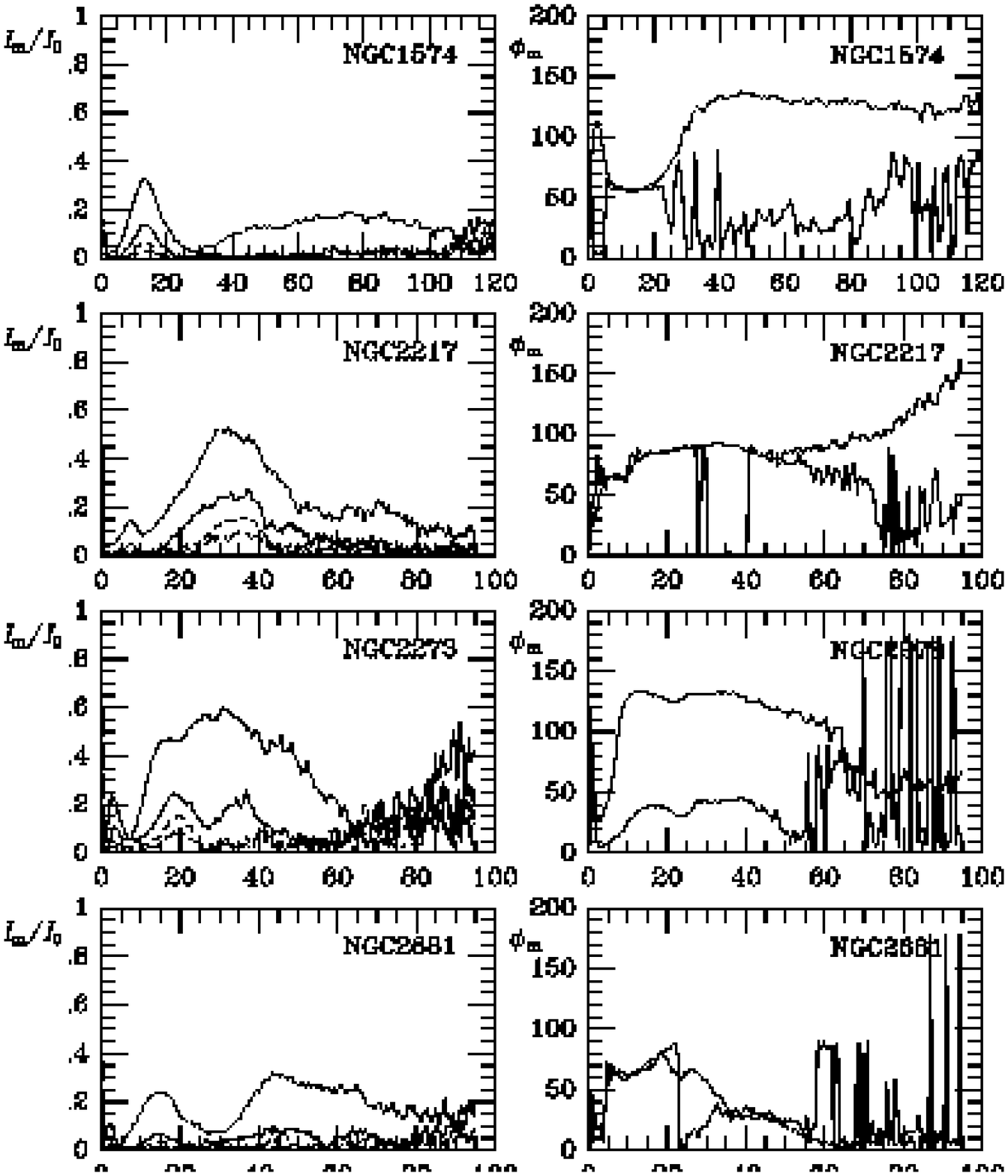}
\caption{}
\end{figure}
\begin{figure}
\figurenum{3 (cont.)}
\plotone{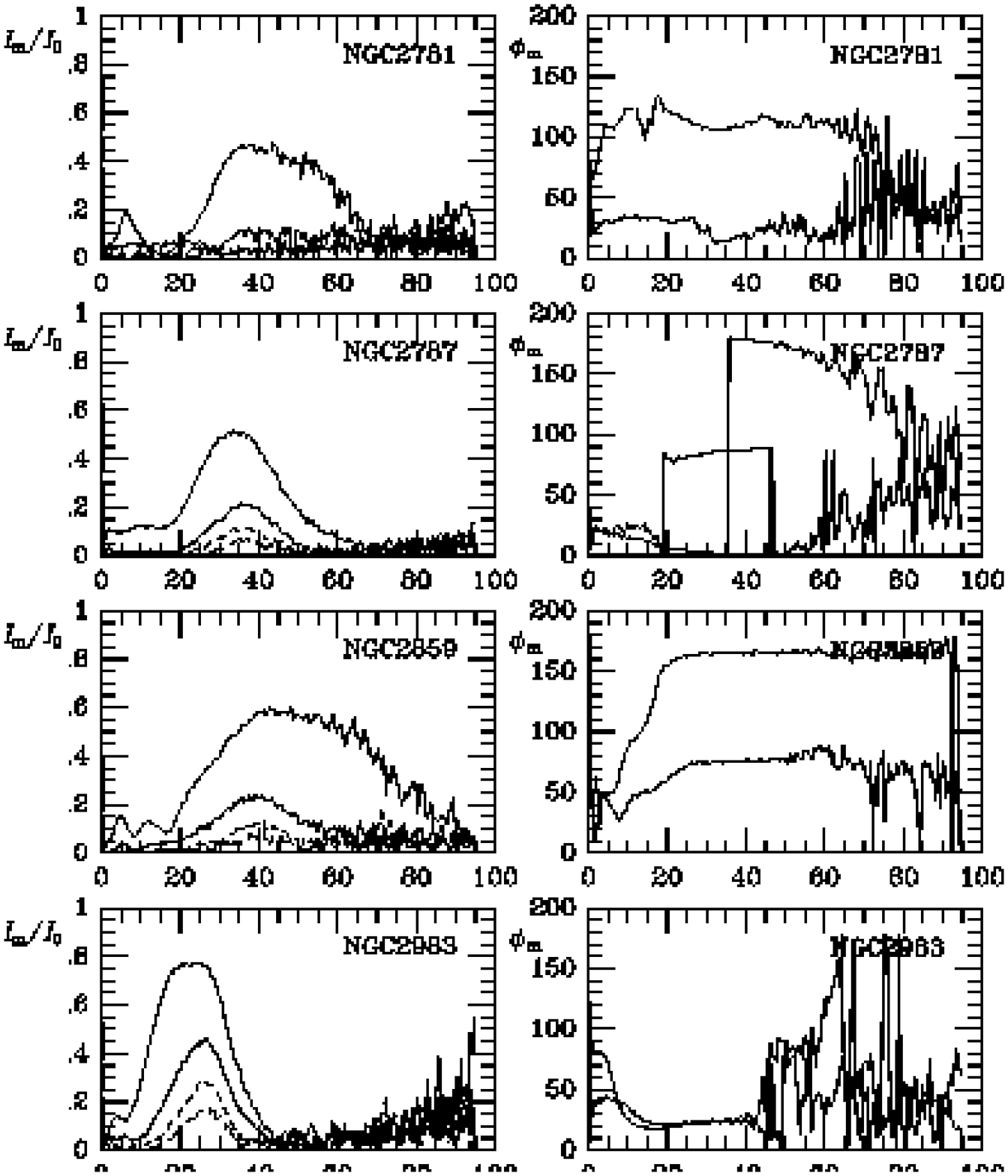}
\caption{}
\end{figure}
\begin{figure}
\figurenum{3 (cont.)}
\plotone{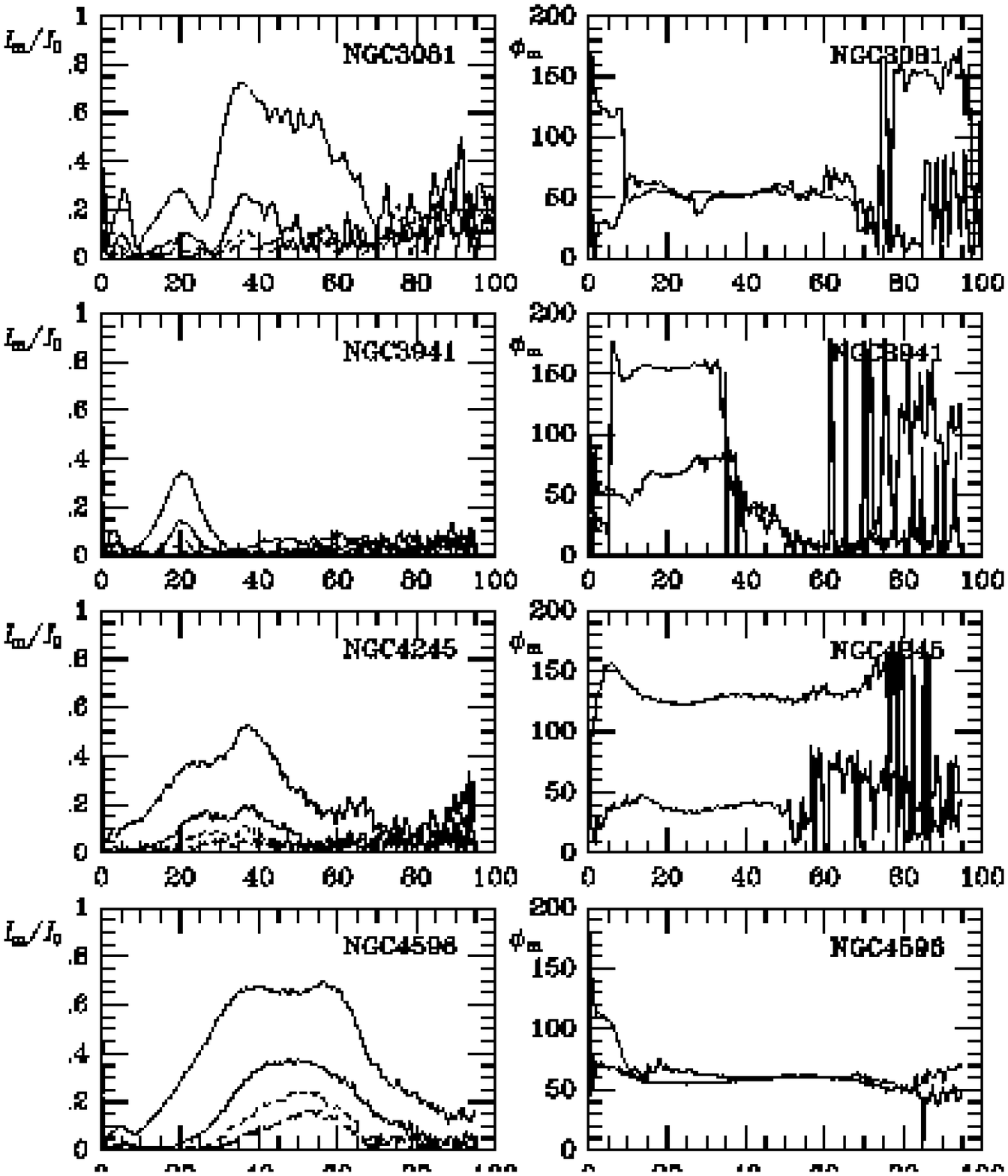}
\caption{}
\end{figure}
\begin{figure}
\figurenum{3 (cont.)}
\plotone{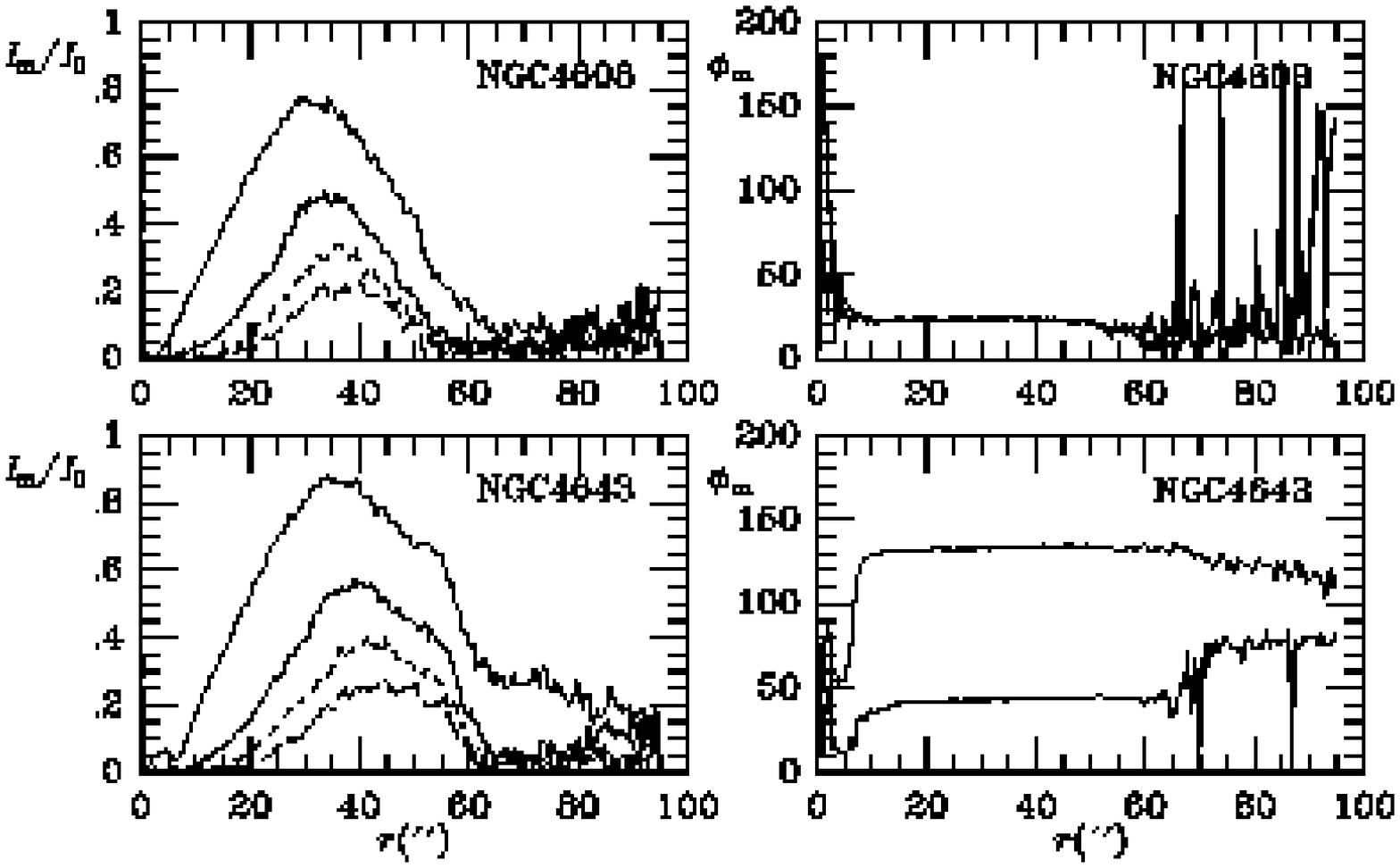}
\caption{Relative Fourier intensity and phase profiles for 26 early-type disk
galaxies. The left panels show the $I_m/I_0$ profiles for $m$=2 (solid curves),
4 (dotted curves), 6 (dashed curves), and 8 (dash-dot curves). The right panels
show the phases $\phi_m$ for $m$=2 (solid curves) and 4 (dotted curves).
The 360$^{\circ}$/$m$ periodicity causes the sharp changes in phase in many
of the plots.}
\end{figure}


\begin{figure}
\figurenum{4}
\plotone{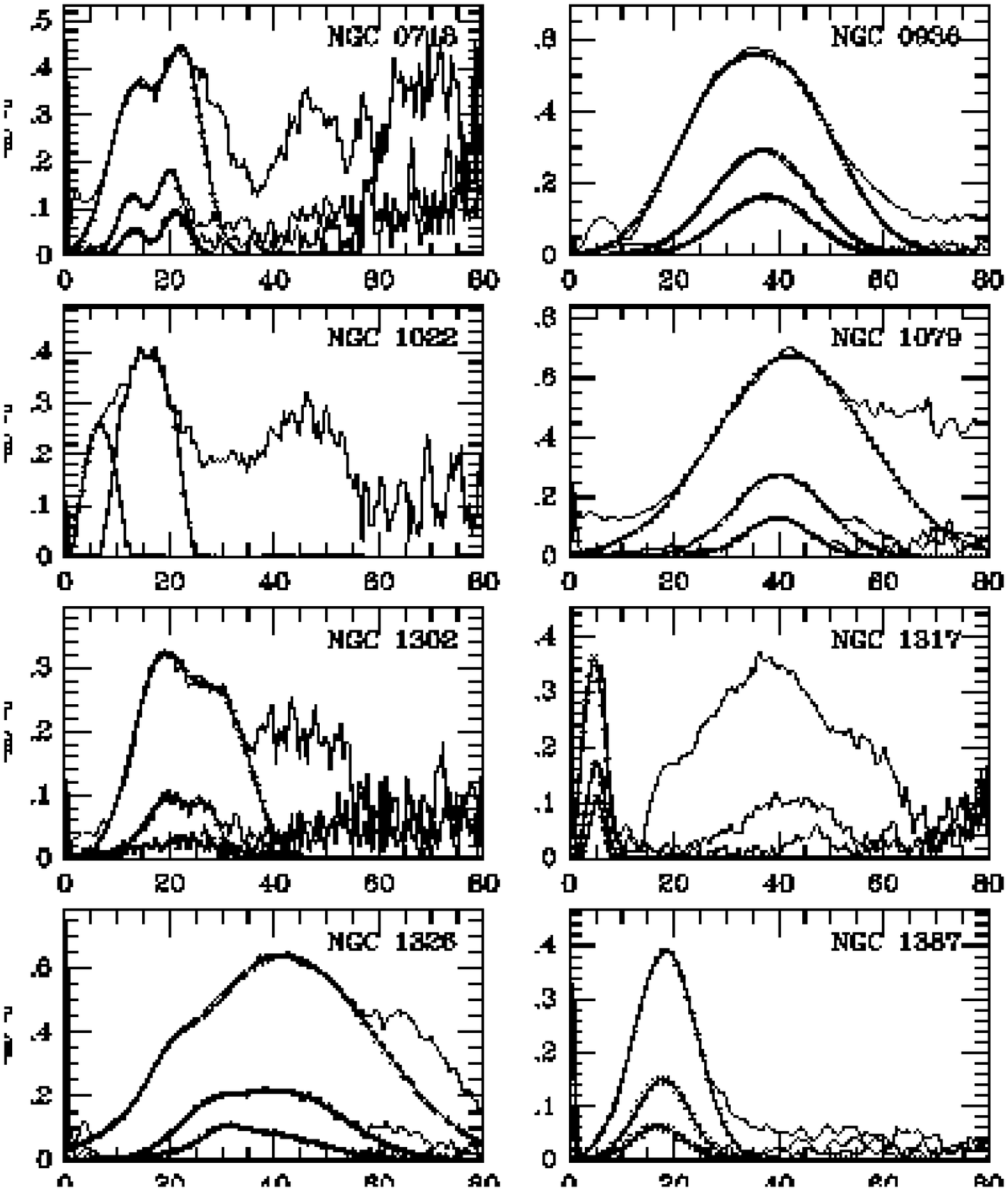}
\caption{}
\label{gaussians}
\end{figure}

\begin{figure}
\figurenum{4 (cont.)}
\plotone{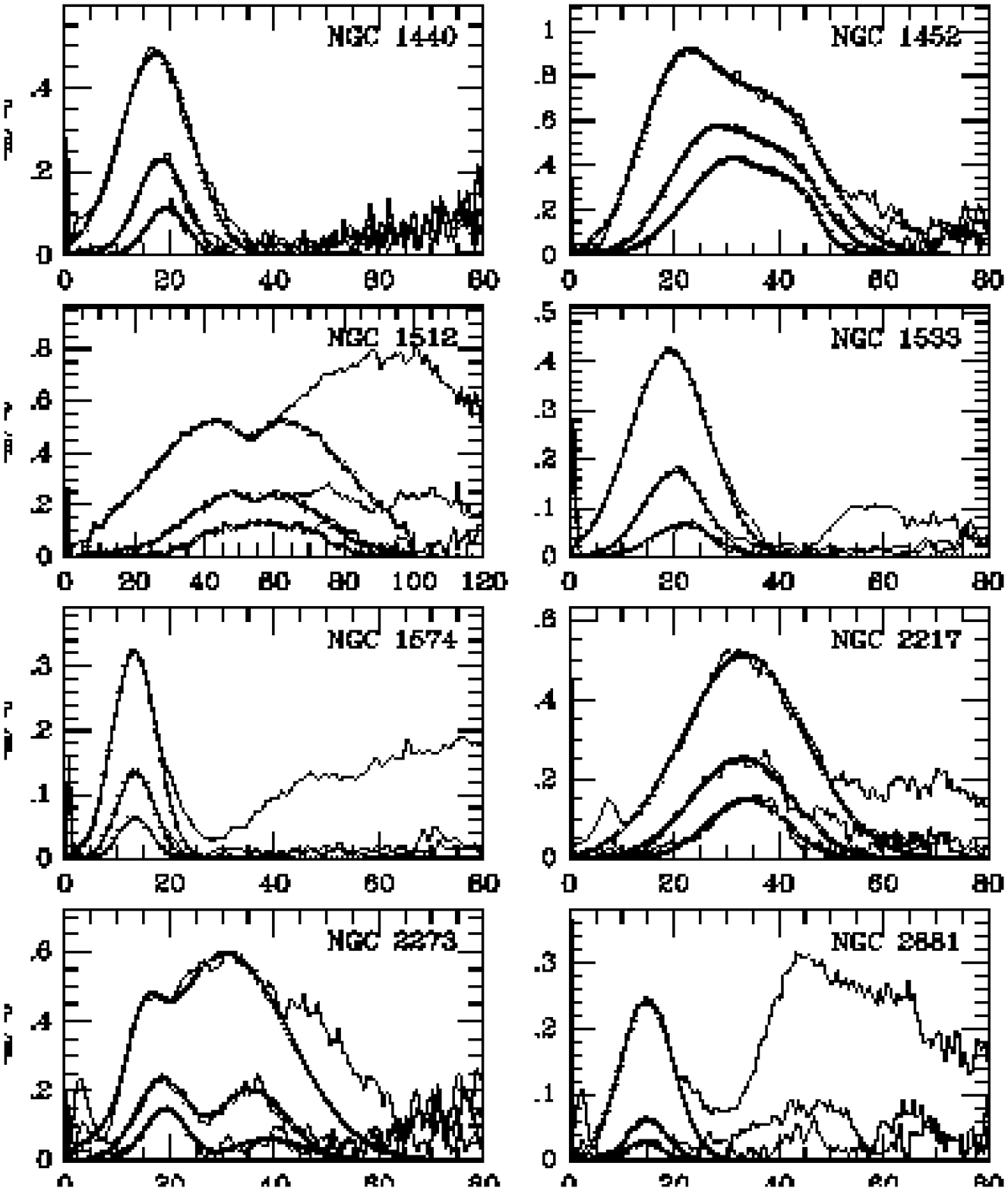}
\caption{}
\end{figure}

\begin{figure}
\figurenum{4 (cont.)}
\plotone{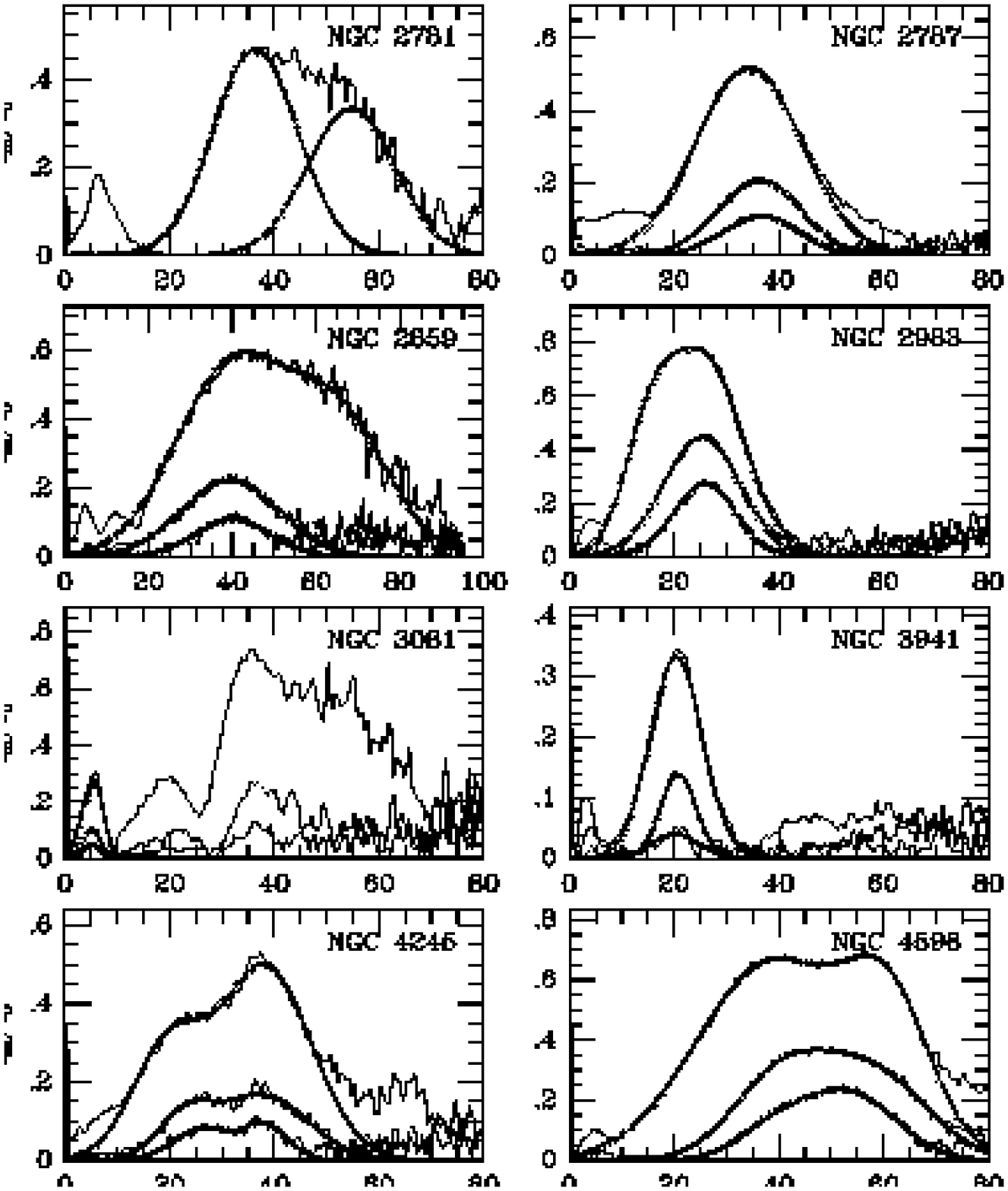}
\caption{}
\end{figure}

\begin{figure}
\figurenum{4 (cont.)}
\plotone{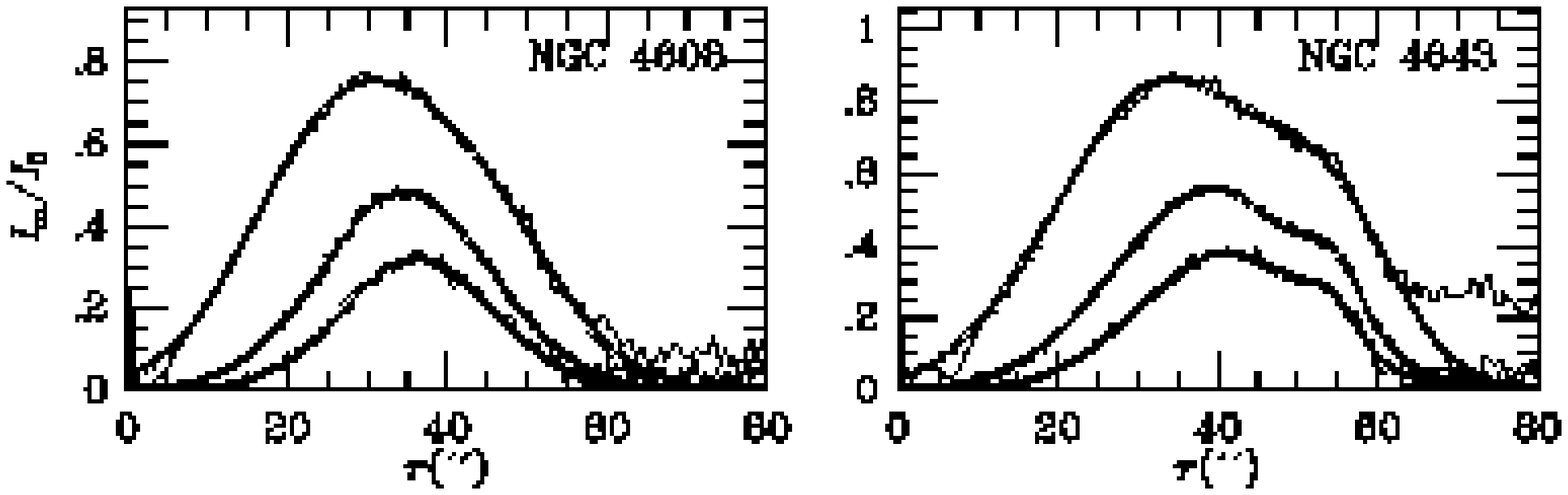}
\caption{Gaussian representations of the $m$=2, 4, and 6 $I_m/I_0$ profiles of
the same galaxies as in Figure 3. Crosses show the fits. In NGC 1022 and 1512,
the crosses are based on the symmetry assumption, not gaussian fits. For NGC
2781, the two gaussians are in slightly different position angles and are
shown separately. Higher order terms than $m$=6 could also be fitted
similarly but are not shown.
}
\end{figure}

\clearpage 

\begin{figure}
\figurenum{5}
\plotone{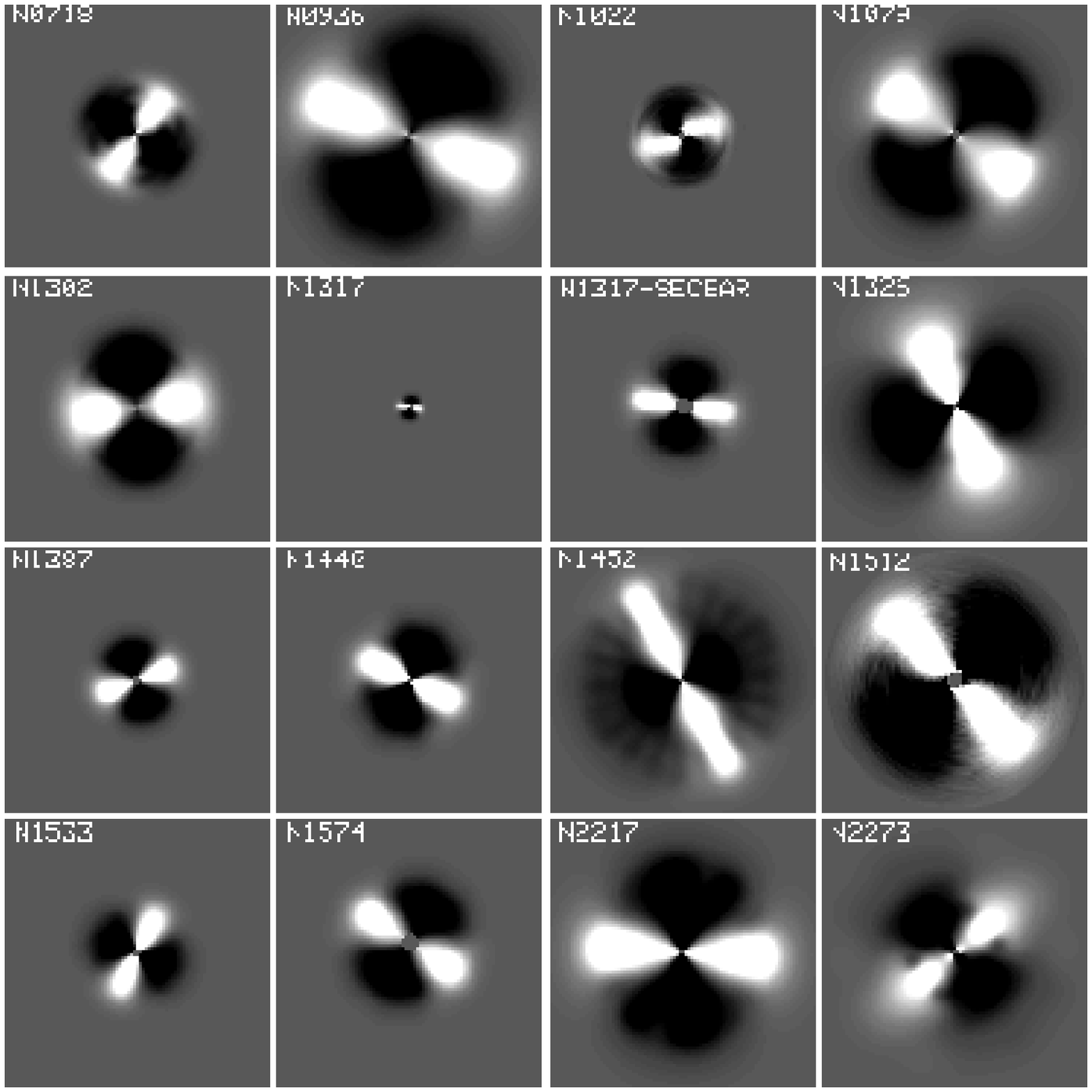}
\caption{}
\label{bars}
\end{figure}
\begin{figure}
\figurenum{5 (cont.)}
\plotone{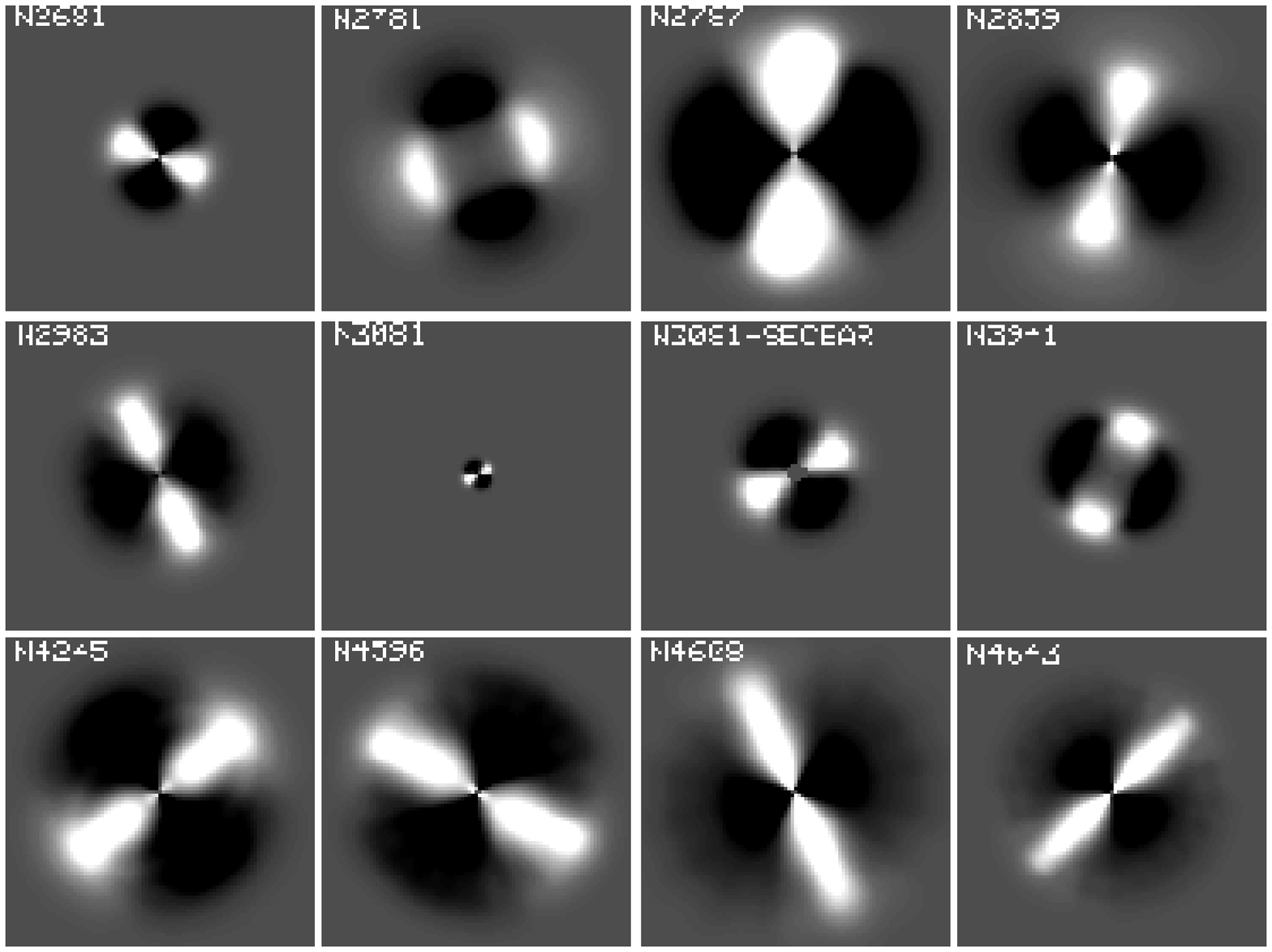}
\caption{Reconstructed images of the bars of the 26 galaxies based on
the gaussian or symmetry assumption representations shown in Figure 4.
Only even Fourier terms up to a maximum $m_{max}$ were used depending on the strength
of the bar. For the strongest bars in the sample, 
$m_{max}$=20. The dimensions of each square are the same as in Figure 2. 
}
\end{figure}

\begin{figure}
\figurenum{6}
\plotone{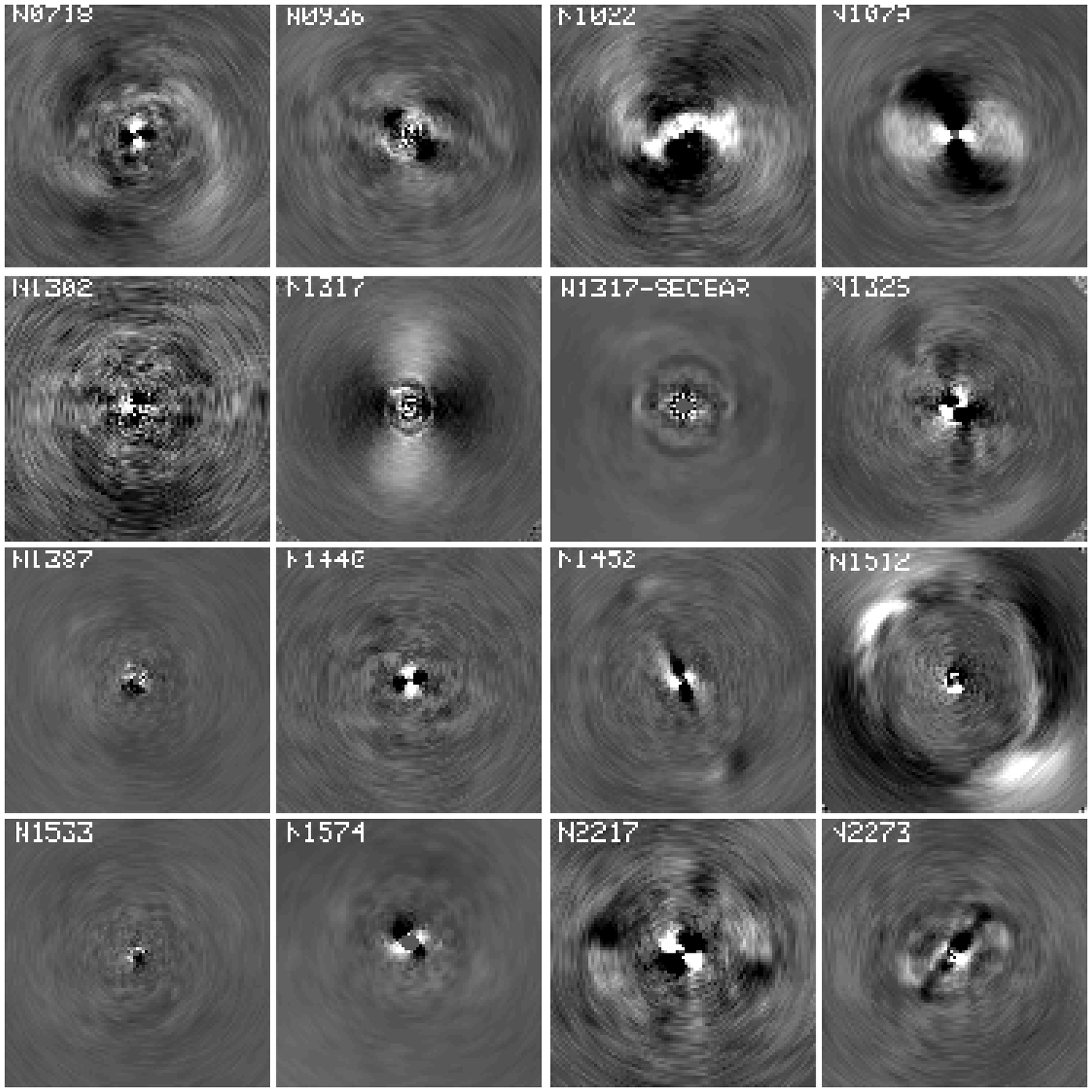}
\caption{}
\label{resids}
\end{figure}
\begin{figure}
\figurenum{6 (cont.)}
\plotone{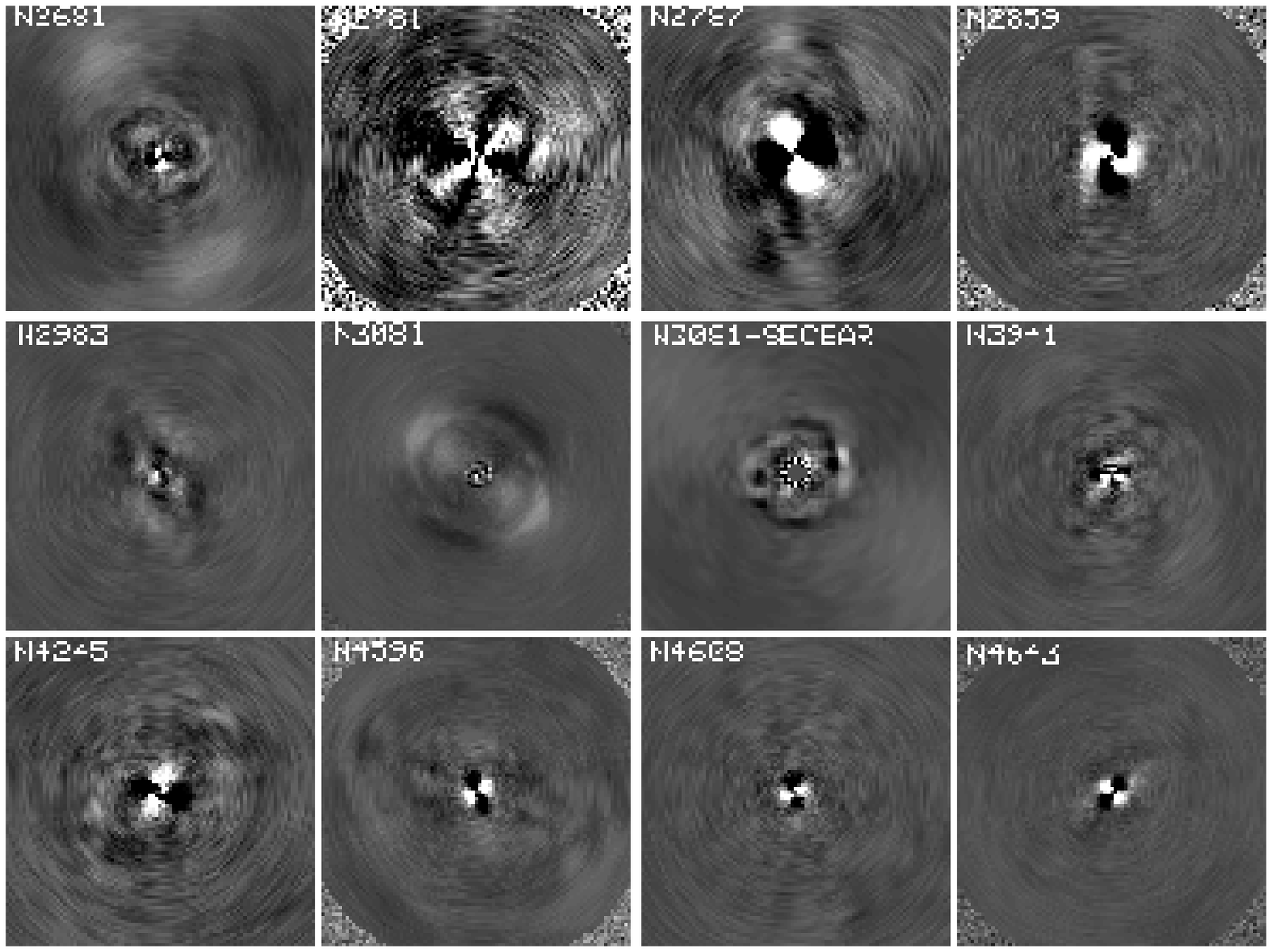}
\caption{Residual $m>$ 0 images of the 26 galaxies on the same scales
as in Figure 5. These include even and odd Fourier terms up to $m_{max}$=20
as needed. These maps show asymmetries, extra bar-like features in the
center, and the imperfections of some of the gaussian fits.
The dimensions of each square are the same as in Figure 2.}
\end{figure}

\begin{figure}
\figurenum{7}
\plotone{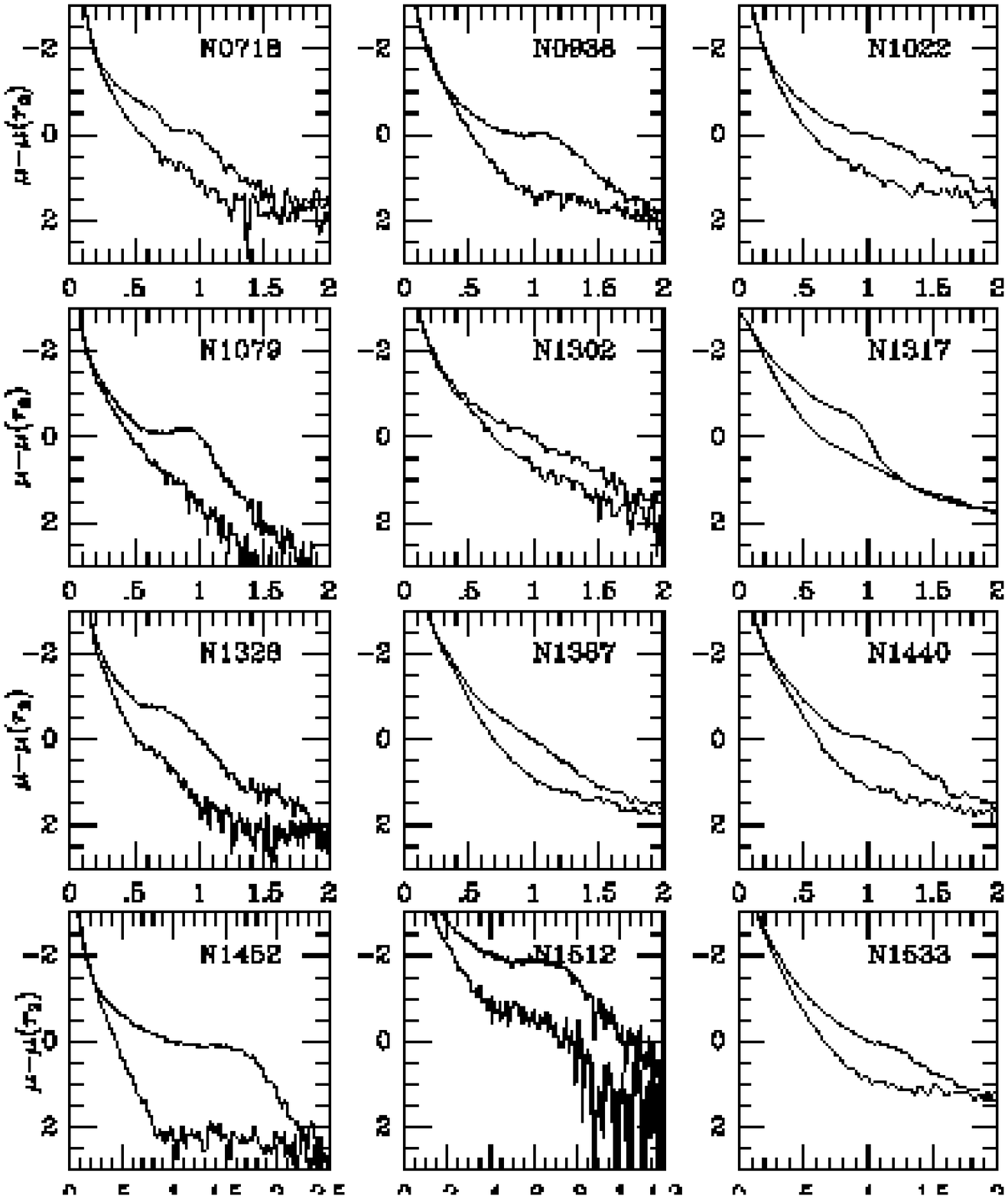}
\caption{}
\label{barprofs}
\end{figure}
\begin{figure}
\figurenum{7}
\plotone{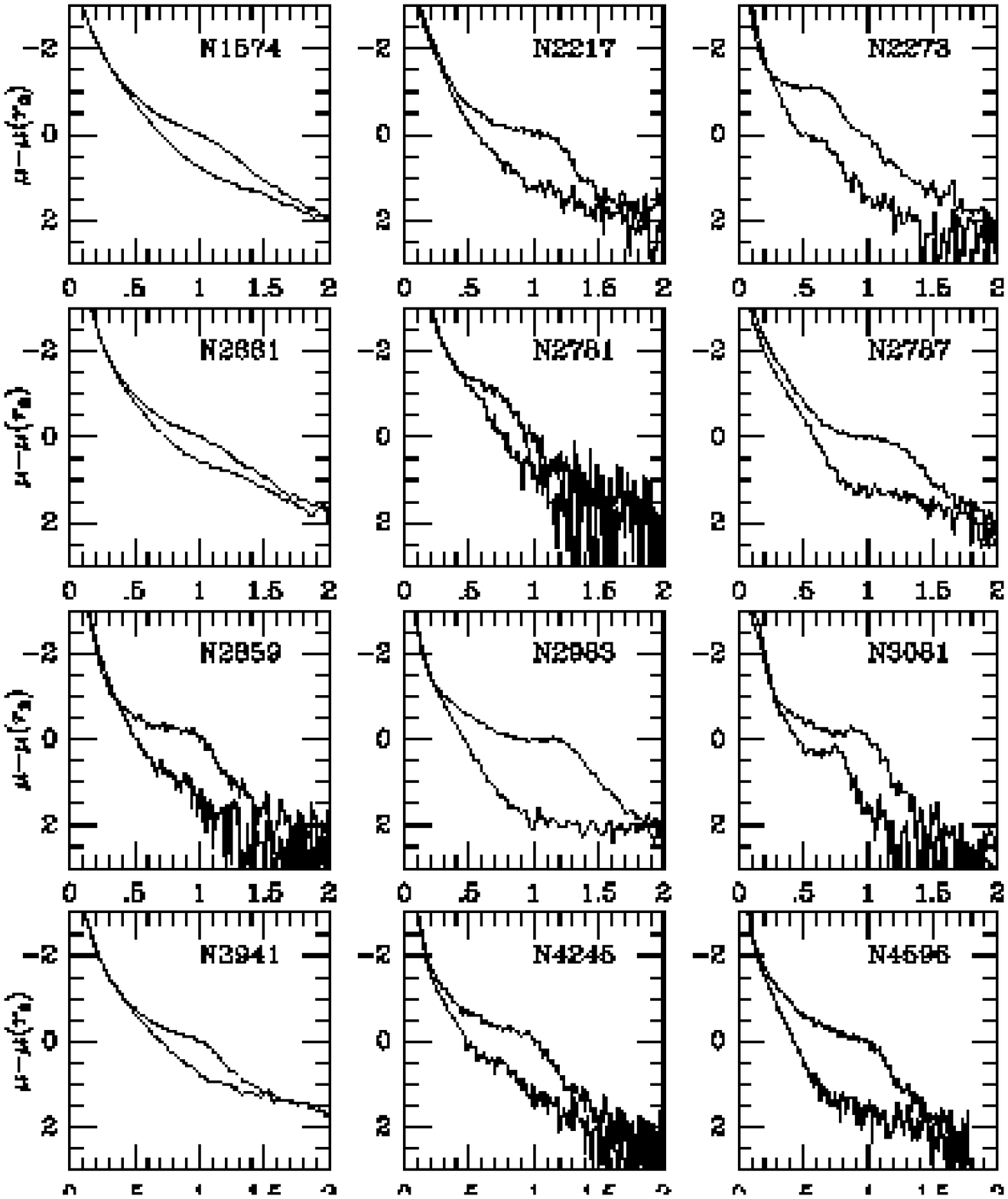}
\caption{}
\end{figure}
\begin{figure}
\figurenum{7 (cont.)}
\plotone{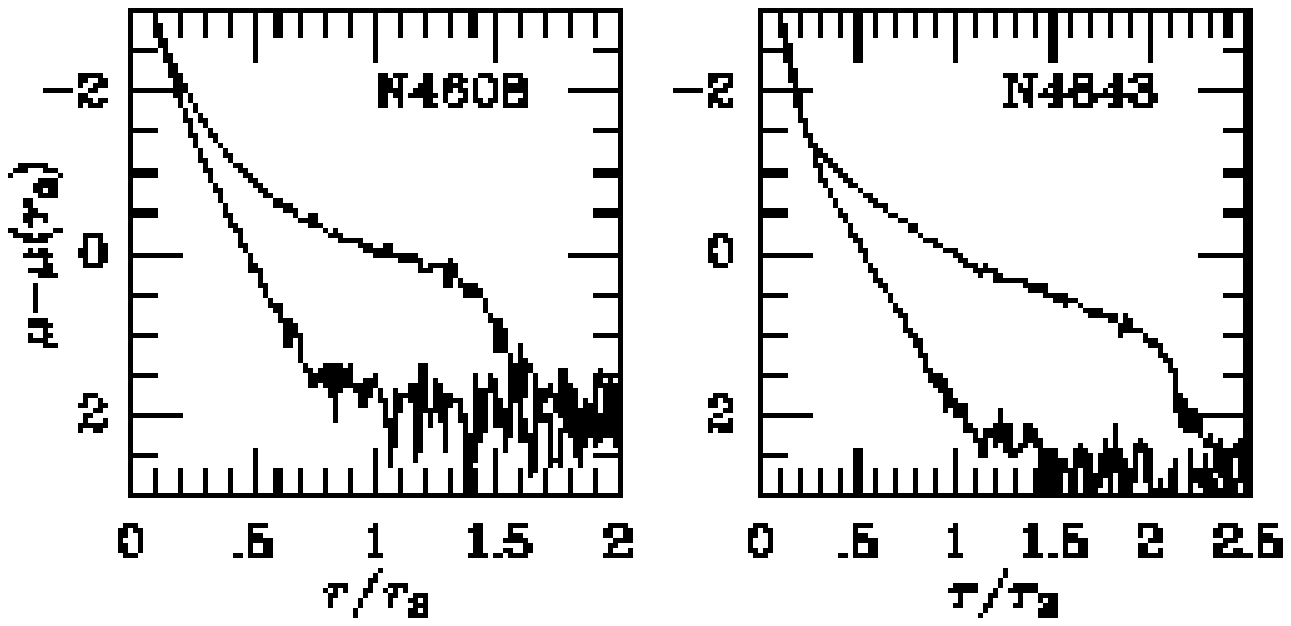}
\caption{Radial surface brightness profiles along the major and minor
axes of the main bar. The radii are normalized to the radius $r_2$ of the
$m$=2 Fourier term, while surface brightness is shown relative to
the surface brightness at this same radius. For NGC 1317, the curves
shown are for the secondary bar, not the primary oval.}
\end{figure}


\begin{figure}
\figurenum{8}
\plotone{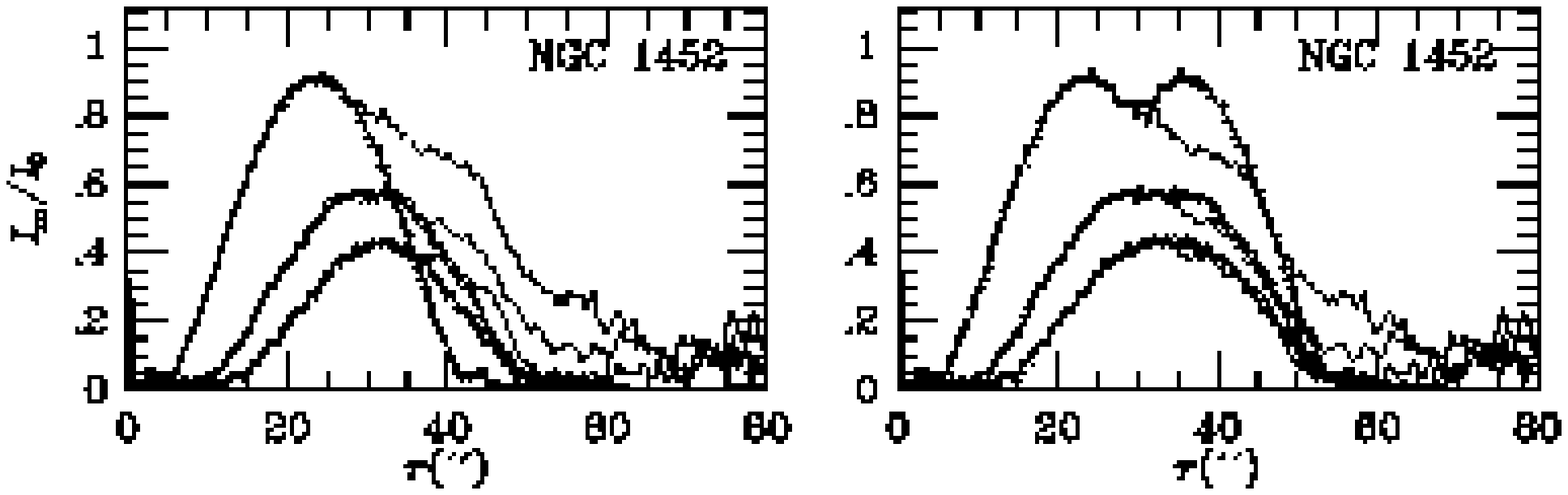}
\caption{Two representations of the bar in NGC 1452 based on application
of the symmetry assumption, as opposed to the gaussian fits shown
in Figure~\ref{gaussians}. The left plot shows the mapping (crosses) for
$r_2$=24$^{\prime\prime}$ while the right plot shows the mapping
for $r_2$=30$^{\prime\prime}$. The latter resembles what we have used
for NGC 1512 in Figure~\ref{gaussians}.}
\label{n1452profs}
\end{figure}


\begin{figure}
\figurenum{9}
\plotone{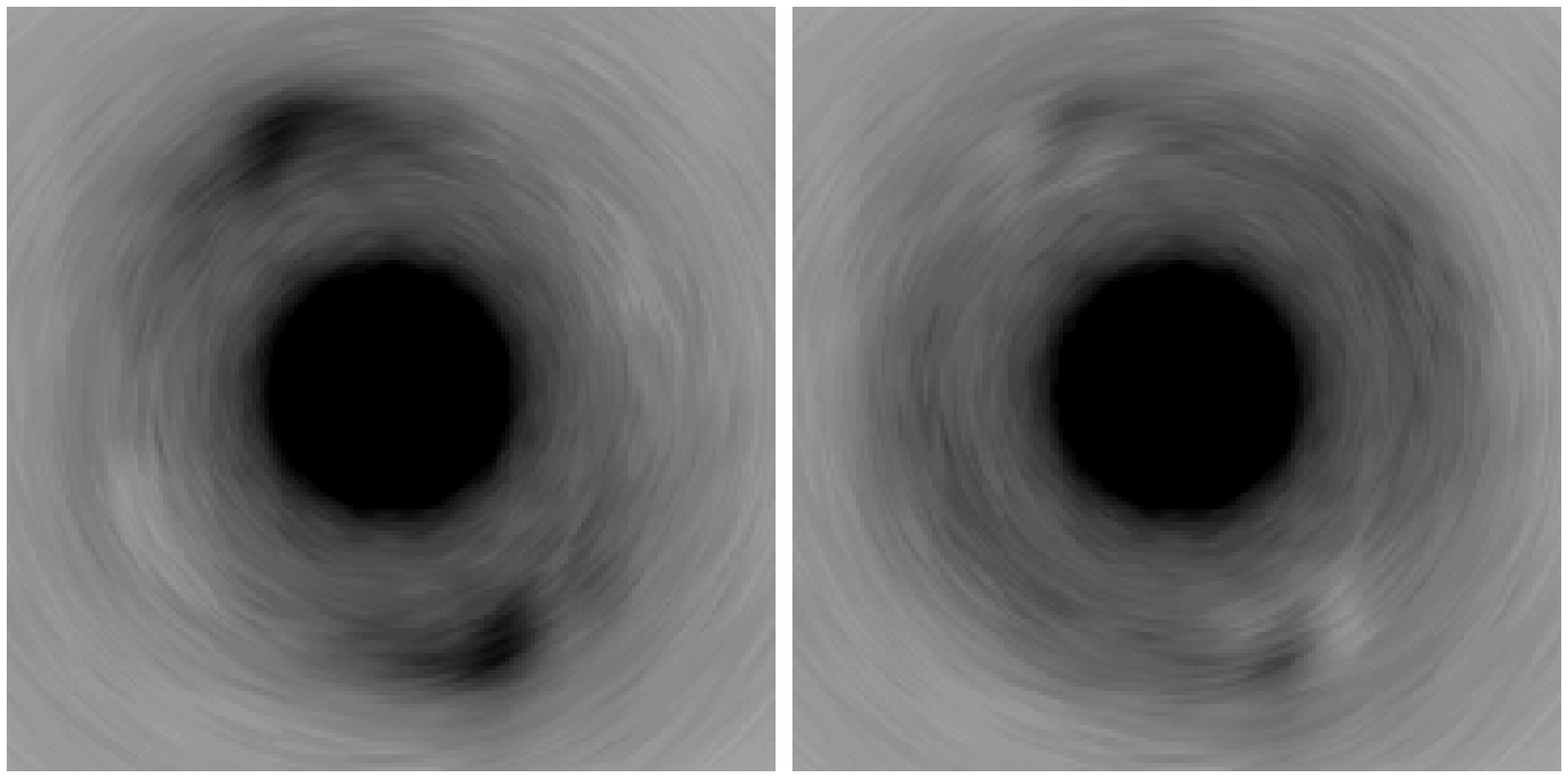}
\caption{Residual intensities in NGC 1452 after removal of the bar
mappings in Figure~\ref{n1452profs}. The left panel is based on
$r_2$=24$^{\prime\prime}$ while the right panel is 
for $r_2$=30$^{\prime\prime}$.}
\label{n1452resids}
\end{figure}


\begin{figure}
\figurenum{10}
\plotone{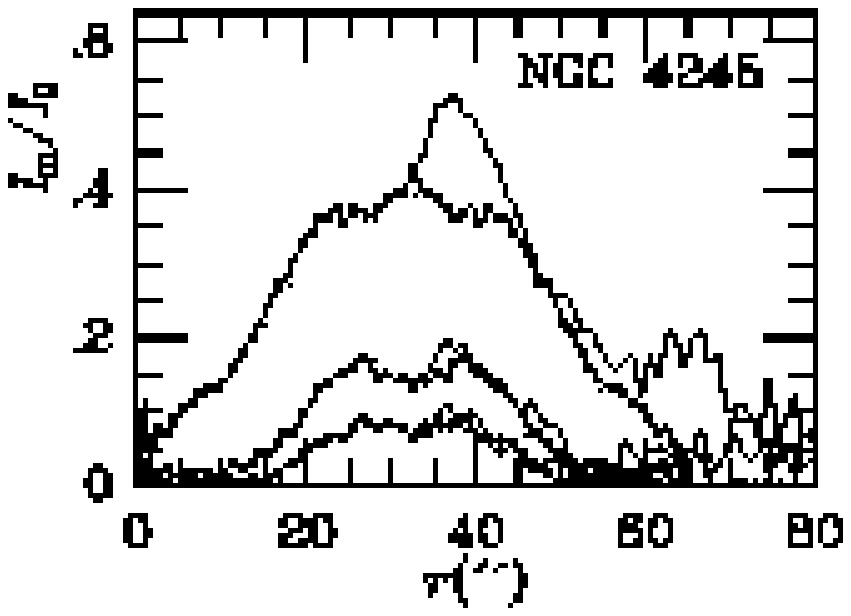}
\caption{A representation of the bar in NGC 4245 based on application
of the symmetry assumption (crosses), as opposed to the gaussian fits shown
in Figure~\ref{gaussians}.}
\label{n4245}
\end{figure}

\end{document}